%% file: skeleton.tex
\documentclass[10pt,conference,letterpaper]{IEEEtran}

\usepackage{mdwmath}
\usepackage{mdwtab}
\usepackage{eqparbox}
\usepackage{subfig}
\usepackage{bm}
\usepackage{fixltx2e}
\usepackage{url}
\usepackage{array}
\usepackage{algorithmic}
\usepackage[cmex10]{amsmath}
\usepackage{amsthm}
\usepackage{cite}
\usepackage[pdftex]{graphicx}
\usepackage{xspace}
\usepackage{color}
\definecolor{comment}{rgb}{0,0.75,0}
\definecolor{keyword}{rgb}{0,0,0.75}
\usepackage{listings}
\usepackage{hyperref}
\usepackage{multirow}

\graphicspath{{figs/}}
\DeclareGraphicsExtensions{.pdf}

\newcommand{\todo}[1]{}

\newtheorem{theorem}{Theorem}
\newtheorem{theorem*}{Theorem}

\input{macros}

\title{Optimal Hierarchical Layouts for \\ Cache-Oblivious Search Trees}

\author{%
{Peter Lindstrom, Deepak Rajan}%
\vspace{1.6mm}\\
\fontsize{10}{10}\selectfont\itshape
Center for Applied and Scientific Computing, Lawrence Livermore National Laboratory\\
7000 East Avenue, Livermore, California 94550, USA\\
\fontsize{9}{9}\selectfont\ttfamily\upshape
pl@llnl.gov\\
rajan3@llnl.gov%
}

\begin{document}

\maketitle

\begin{abstract}
\input{abstract}

\end{abstract}

\todo{This file is a placeholder for the flow of the paper.}

\section{Introduction}

\todo{
Talk about searches, and how B-Trees optimize a two-level hierarchy.
Hard to generalize to multiple levels.
Introduce Cache-Oblivious B-trees. Review papers, and mention how they optimize for all levels.
Mention generalization (Streaming, string, etc.), but *all build upon COBT.*
}

\note{Talk about the effect of caches, and how/why data has to be optimized for the cache. Done.}

\note{Need to reference Safro's MLogA papers. Done.}

In today's computer architectures, the memory hierarchy is becoming increasingly complex, both in terms of number of levels and in terms of the difference in performance from one level to the next.
As a result, algorithms and data structures that are designed for flat (or even two-level) memory with uniform access times can result in significantly suboptimal performance.
In this paper, we are interested in improving memory access locality for search trees via data reordering.
The classic search tree is the B-tree~\cite{Bayer:Btrees}, which has been designed for a two-level cache hierarchy, and is usually optimized for a particular block transfer size (\eg, a cache line or disk block). It is not clear if B-trees can be successfully optimized for a multi-level cache hierarchy, with one level per transfer block size. Furthermore, B-trees are known to perform poorly when the \nodes of the search trees are of different sizes (\eg, when the search keys are variable-length)~\cite{Bender:COStringBT}. As a result, cache-oblivious search trees have been suggested in the literature. In this paper, we present a new locality measure that can be used to derive cache-oblivious data structures.
Focusing our attention on search trees, we show how optimizing our locality measure results in better cache-oblivious search tree layouts than prior layouts.

The fundamental structure commonly employed for cache-oblivious search trees is the \vEBl. First introduced by Prokop~\cite{Prokop}, these recursively defined layouts are similar to van Emde Boas trees, hence the name. 
In~\cite{Brodal:COSearch-vEBLayout}, this layout was shown to result in much better binary search times than simpler orderings such as breadth-first and depth-first pre- and in-order.
\comment{
, and has been used (with minor variations) ever since to design cache-oblivious search trees.
}
Minor variants of these \vEBls have since been used in a variety of other settings. 
In~\cite{Bender05}, the authors introduce a very similar layout that differs only in how the tree is partitioned at each branch of the recursion. Using this layout as the basic building block, they present dynamic search trees, and refer to these as cache-oblivious B-trees. In~\cite{Bender:COSearchCost}, the authors provide bounds on the asymptotic cost of cache-oblivious searching. By analyzing a generalized version of the \vEBl, they provide a modified version that is arbitrarily close to the asymptotic bound. 
In~\cite{Bender:COStringBT,Bender:BTDiffKeys}, the authors address the problem of building cache-oblivious layouts of search trees with variable-sized search keys. They use a modified version of the \vEBl in which the tree is partitioned differently. In~\cite{Bender:COStreamingBT}, the authors present two cache-oblivious streaming B-trees -- data structures that implement cache-oblivious search trees optimized for dynamic insertions and deletions. Again, these rely on a version of the \vEBl with a slightly different partitioning scheme.
\comment{
An alternate approach for handling multi-level caches is to design algorithms rather than data structures that are cache-oblivious. In \cite{Frigo:COAlgos}, the authors present asymptotically optimal cache-oblivious algorithms for many commonly used subroutines such as sorting and fast Fourier transforms. Building upon previous work, Brodal and Fagerberg~\cite{Brodal:CODisSweep} present cache-oblivious versions for many divide-and-conquer algorithms.
}
We note that cache-oblivious data structures are not limited to search trees. They have been proposed in a variety of settings, some of which include hash tables \cite{Pagh:COHash}, meshes \cite{Yoon05,Bender:COMesh}, and Bloom filters \cite{Bender:COBloomFilter}.

\subsection{Contribution: Cache-oblivious Hierarchical Layouts}

We describe a general framework for generating search tree layouts, and present new orderings from this framework that result in better cache-oblivious search tree layouts than those suggested in the literature. We refer to all layouts that fit the new framework as \emph{Hierarchical Layouts}. 

\comment{Furthermore, major improvements can often be obtained by considering simple-to-describe modifications to the layouts used currently, which we show are special cases of a general framework for generating layouts. We refer to all layouts that fit the general framework as \emph{Hierarchical Layouts}. 
}

Consider a tree $T$ of height $h$, \ie, with $h$ levels of \nodes. For ease of exposition, we restrict our discussion to complete binary trees;  therefore the number of \nodes is $2^h-1$. 
\comment{We can think of any \node in this tree as belonging to a particular level, where}
Counting the levels from top to bottom, the root is on level $0$ and the leaves are on level $h-1$.
Observe that level $i$ has $2^i$ \nodes. 

Any \HL can be described recursively as follows:
Partition $T$ by cutting it horizontally between level $g-1$ and $g$, which results in a top subtree $A$ of height $g$ with
$2^{g-1}$ leaves.
Let $L_A$ be the set of leaves in the top subtree $A$.
Given a leaf \node $x$ in $L_A$, we say that a bottom subtree formed by a child $c$ of $x$ and the descendants of $c$ is a \textit{child subtree} of $x$.
\comment{Since there are }
With $2$ child subtrees for each \node in $L_A$, we have $2^g$ bottom subtrees 
of height $h-g$.
Any relative ordering of the recursive subtrees that arranges them consecutively in memory constitutes a \HL. 
The \node ordering within each subtree is given by recursive application of this
decomposition, until each subtree consists of a single \node.
\comment{
To ensure that the exact position of top subtree $A$'s root vertex is consistent with its ordering relative to the bottom subtrees (in- or pre-order), all the subtrees recursively obtained as top subtrees in subsequent cuts of top subtree $A$ must be arranged in the same fashion.
}

The effectiveness of any particular \HL as a cache-oblivious search tree depends on the relative ordering of subtrees and on the height of the top subtree $g$. In this paper, we focus on finding optimal cuts and orderings of the subtrees to maximize locality and minimize cache misses, and propose a new cache-oblivious \HL.

We also show that the widely used \vEBls are a
special case of \HLs; therefore any cache-oblivious search tree data structure that utilizes a \vEBl can be improved by switching to our proposed layout. The main take-home message of this paper is that the widely used version of the \vEBl is {\bf not} the best \HL. Significantly better cache-oblivious layouts can be obtained by considering \HLs that minimize our measure.

\comment{describe some notation that allows us to characterize all \HLs, and}
In \autoref{sec:vebl}, we 
motivate some simple improvements to the \vEBl. \autoref{sec:wep} describes a mathematical measure of locality for tree orderings that correlates well with cache-miss ratios, thus resulting in cache-oblivious layouts. We analyze which \HLs perform better with respect to this new measure, the \emph{Weighted Edge Product}. We improve the layout further in \autoref{sec:minwep} by deriving \minwep, the \HL that minimizes the \WEP. Our experiments indicate that \minwep on average improves performance by almost 20\% compared to the layouts described in the literature.

\subsection{\HLs: A Nomenclature}

\comment{
Recall that 
}
A \HL is
given entirely by (1)~the height at which the tree is partitioned, 
(2)~the position of the top subtree relative to the bottom subtrees, and (3)~the relative ordering of the bottom subtrees.  
\note{We can't represent all \HLs using our notation (only pre and in). \HLs restricted.}
This definition allows for a very large combination of cut heights and orderings.
\comment{many of which do not yield cache-oblivious layouts.}
For this reason, we impose additional restrictions; the motivation behind some of them will become clear later in the paper. We refer to layouts belonging to this restricted set as \RLs because they can be 
categorized entirely using a small set of recursive rules and parameters, allowing for a more compact nomenclature
than the more general \HL.
\comment{
would be possible for 
}

In a \RL, at any branch of the recursion that cuts a subtree into its top subtree $A$ and the corresponding bottom subtrees, we enforce the following restrictions. 
(a) $A$ is arranged either in the middle of all the bottom subtrees (in-order), or at one end (pre-order). 
(b) The top subtree obtained in the partitioning of $A$ must be arranged relative to the bottom subtrees in the same fashion as $A$. (c) If $A$ is arranged in-order, we choose the children of the leftmost $2^{g-2}$ leaves in $L_A$ to be the bottom subtrees on the left of the top subtree. (d) If 
any bottom subtree is arranged in-order, 
all bottom subtrees that are arranged further away from $A$ are also arranged in-order. (e) 
Looking outwards from $A$, the bottom subtrees
are ordered in the same order as that of the parent leaves $L_A$, or in the reverse order.  
(f) If $A$ is arranged pre-order, then it is placed on the side of the bottom subtrees that is closer to its parent leaf. Thus, we 
use pre-order layouts to refer to both pre-order and post-order arrangements of the top subtree, depending upon the context. (g) The cut height $g$ is a function only of the height of the subtree.

Based on the preceding discussion, we present a new nomenclature for categorizing \RLs, an important subset of \HLs. A \RL is categorized as $\PR$ for pre-order and $\IN$ for in-order to indicate the arrangement for the outermost branch of the recursion (when we cut the tree $T$ itself). At each branch of the recursion, the position of the first in-order bottom subtree, counting outwards from the top subtree, is indicated as a subscript. If all the bottom subtrees are arranged pre-order, then we denote this by $\infty$. The cut height $g$ (as a function of the height of the subtree $h$) is indicated as a superscript. 
We indicate an layout where the bottom subtrees are arranged in reverse order of the leaves $L_A$ using the $\sim$ symbol on top.

Bringing this all together, we see that $\alt{\IN}^1_\infty$ is the \RL that always cuts at height $1$, arranges the top subtree in the outermost branch of the recursion in-order, and arranges all other subtrees pre-order in the reverse order of the top subtree leaves. In \autoref{tbl:notation}, we categorize all the layouts we consider in this paper using this nomenclature. All the layouts described in this paper belong to the restricted set of \RLs.

\comment{
For pre-order layouts, it can easily be shown that the top subtree should be arranged at the end of the bottom subtrees that is closest to the parent leaf of the top subtree at the previous branch of the recursion.
}
\note{I don't understand the sentence ``For pre-order layouts...''  The relationship is too complicated: top subtree should be arranged at A of B that is C to D of E at F of G. Made clearer, see (e)}
\comment{Since this could potentially be on either side,}
\note{Either side of what? Fixed.}
\comment{
we use pre-order layouts to refer to both pre-order and post-order arrangments of the top subtree, depending upon the context.
}
\note{This is important but far from clear.  Why are we saying in the intro where subtrees should be?  This seems too early---we haven't really described the problem yet, even at a high level. After the contributions, and part of the restriction description, for now.}
\comment{
On the left side of an in-order top subtree, the left-to-right ordering of the bottom subtrees is in fact post-order. 
Since this is just a mirror image of the pre-order arrangement on the right side of the top subtree, for ease of exposition, pre-order is used to refer to both pre- and post-order. 
}

\comment{
For most of this paper, we set $g = \FLOOR{h/2}$. In \autoref{sec:cut-hts}, we will revisit the question of the best cut height.
}

\note{I would expect this section to explain what hierarchical layouts are, not how to optimize them (wrt. some unknown criterion).  We're only in the intro. Not explained, just stated, after motivation.}

\todo{
Goal of paper: vEB layouts use the wrong layout. Outline: Minor modification (pre to in vastly better.)
Insight from why in better guides the metric of choice (WEP). WEP is lower for in. Can we do better?
Our layout - combination of pre and in.
}

\section{Cache-oblivious \HLs}
\label{sec:vebl}

In this Section, we motivate better cache-oblivious orderings within the framework of \HLs. First, we review the \vEBls used in the literature, which are a special case of \HLs.
\comment{and are broadly referred to as \vEBls.}
In Prokop's ordering~\cite{Prokop}, the subtrees are cut at height $g = \FLOOR{h/2}$, the top subtree is placed before the bottom subtrees, and then this ordering strategy is applied recursively to each subtree. 
The bottom subtrees 
\comment{($B_1, \ldots, B_{2^g}$)}
are arranged in the same order as the order of their parent leaves $L_A$. 
\comment{
In this ordering, the subtrees are ordered as $A, \bm{B_1}, \bm{\ldots}, \bm{B_{2^g}}$, and then this ordering strategy is applied recursively to each subtree. 
}
For the rest of this paper, we refer to this version of the \vEBl as the pre-order \vEBl, and denote it as \pvl. In our nomenclature, \pvl is $\PR^{\FLOOR{h/2}}_\infty$ (see \autoref{tbl:notation}).
\comment{A bottom subtree in boldface indicates that its \nodes are arranged in pre-order.}

\autoref{fig:tree-vebo} illustrates \pvl for a tree of height $6$.
\comment{using $g = \FLOOR{h/2}$.} The number inside each \node is its position in the final layout, ranging from $1$ to $63$. 
Observe that at every branch of the recursion, the top subtree is arranged pre-order. In the outermost branch of the recursion, the \nodes in the top three levels are arranged first (positions $1$ to $7$).
\autoref{fig:tree-layouts} also indicates the length of each edge, \ie, the difference in position of its \nodes, using lines whose thickness is inversely proportional to the length. 

In Bender's layout~\cite{Bender05}, the authors set $g = h - 2^{\CEIL{\log_2(h/2)}}$. In other words, the height of the bottom subtrees is set to the largest power of two smaller than $h$. The authors refer to their layout as a \vEBl since it is similar to the layout introduced in~\cite{Prokop}. Nevertheless, we make the distinction that only \HLs that set the cut height $g=\FLOOR{h/2}$ are \vEBls. \bender is identical to \pvl for trees whose height is a power of two.
For all other tree heights, \bender layouts have smaller top subtrees, compared to \pvl. \autoref{fig:tree-bender} illustrates Bender's layout. Observe that the \nodes in the top $2$ levels are arranged next to each other, indicating a cut height of $2$ at the outermost branch of the recursion. In our nomenclature, \bender is $\PR^{h - 2^{\CEIL{\log_2(h/2)}}}_\infty$ (see \autoref{tbl:notation}).
\comment{see that \bender is in the same column as \pvl, but is not a \vEBl, and is therefore placed in its own row.}

We will see later that \HLs also include all the simple and commonly used layouts such as in-order, pre-order, and breadth-first. One can think of cut heights $g=1$ and $g=h-1$ as the extreme cases, corresponding to these simple layouts. We will also show that cache-oblivious layouts are obtained by cutting the tree near the center, with $g$ approximately equal to $h/2$. 

\note{Move next 2 paragraphs to introduction, merging it with text there. Done.}

\subsection{Evaluating layouts using block transitions}
\label{sec:blocktrans}

To compare \HLs, we will estimate the number of cache misses for a particular
cache block size and layout as follows. Consider a cache consisting of a single block that can hold \csize
data elements, and which is backed by a larger memory consisting of several
such blocks.  (In practice caches tend to hold more than one block,
but that would unnecessarily complicate our derivation.)
Let $i$ and $j$ be data elements stored in blocks $B(i)$ and $B(j)$, respectively,
and let $\len[ij]$ denote the difference in position of $i$ and $j$ on linear
storage.
For ease of exposition, we set $\len[ij] = \len$. 
Suppose $i$ is accessed first, bringing $B(i)$ into the cache.
We wish to estimate the probability of a cache miss when $j$ is accessed next.
Clearly, if $\len \geq \csize$, then a cache miss is inevitable, since then
$i$ and $j$ are stored in different blocks.
When $\len < \csize$, the likelihood of a cache miss
depends on the positions of $i$ and $j$ within their blocks.
In absence of further information, we will assume that the position
of $i$ within $B(i)$ is distributed uniformly, and similarly for $j$.
(Even in
practice, modern operating systems allocate memory blocks with
nearly arbitrary alignment.)
Hence, there are $\ell$ out of $\csize$ possible alignments
that separate $i$ and $j$ into different blocks, and the
probability of a cache miss occurring when $j$ is accessed is therefore
\begin{equation}
\label{eq:prob-cmiss}
  \acmr[\csize][\len] =
  \begin{cases}
        \frac{\len}{\csize} & \text{if $\len \leq \csize$} \\
        1 & \text{otherwise}
  \end{cases}
\end{equation}

To represent a particular access pattern on the data, we use the notion of an \emph{affinity graph}, as in~\cite{Yoon05,Yoon06}.
We model the data elements as \nodes \nodeset in a graph $G(\nodeset, \edges)$, with an undirected edge
indicating a nonzero likelihood that its two \nodes be accessed in
succession.
The affinity between $i$ and $j$ may be expressed in terms of a
weight $\weight[ij] = \weight[ji] > 0$.  
\note{We also use $A$ to denote the top subtree. Changed to \delta. Changed back to A.}
Let $A$ denote the matrix of affinities, such that $a_{ij} = w_{ij}$ if $ij \in E$
and $a_{ij} = 0$ otherwise. We model data accesses as a Markov
chain random walk on $G$ with transition matrix $P = D^{-1} A$,
where $D$ is the diagonal matrix with $d_{ii} = \sum_j a_{ij}$.
If $G$ is strongly connected, as is the case for binary trees,
then it is well-known that the probability $\Pr(X_t = i, X_{t+1} = j)$
of being in state $i$ and transitioning to state $j$ equals
$\frac{\weight[ij]}{W}$, where $W = \sum_{ij \in E} \weight[ij]$.
In other words, the probability of accessing two data elements in
succession is proportional to the weight of the edge connecting them.

In a binary search tree $T$, the affinity graph is the search tree itself, and the search for a particular element results in a walk from the root on level $d = 0$ of the tree to the \node representing the element. 
Therefore, only the \node searched for and its ancestors are visited, beginning with the root, and
thus \nodes near the top of the tree are more likely to be
visited than \nodes near the bottom. Assuming each \node is equally likely to be searched for, the likelihood of traversing
a given edge between levels $d-1$ and $d$ in a tree of height
$h$ is
\begin{equation}
  \prob[d,h] = \frac{\nodeset[h-d]}{\nodeset[h]} = \frac{2^{h-d}-1}{2^h-1},
  \label{eq:edgewts}
\end{equation}
where $\nodeset[h]$ is the number of \nodes in a complete binary tree of height $h$.
For an edge $ij$ between levels $d-1$ and $d$, we set $\weight[ij] = \prob[d,h]$, 
which ensures that the probability of accessing two data elements in succession is proportional to the weight of the edge connecting them. Near the top of the tree, where $2^h > 2^{h-d} \gg 1$, this
likelihood decreases approximately geometrically by level,
\ie $\prob[d,h] \approx \prob[d] = 2^{-d}$.  We will
use these approximate probabilities and corresponding edge
weights for the rest of this paper, primarily for ease of
analysis.

Given this probability of accessing any two \nodes in succession, the expected percentage of consecutive accesses
that will result in a cache miss for a particular block size \csize is
\comment{ given by}
\begin{equation}
  \PB(\csize) = \frac{1}{W} \sum_{ij \in E} \weight[ij] \acmr[\csize][{\len[ij]}]
\end{equation}
For any given layout and block size, we refer to \PB as the \emph{Percentage of Block Transitions}.
If one layout dominates another for all block sizes under this metric, then clearly it will result in a better cache-oblivious layout. For a particular block size, we can also calculate \PB for all tree heights $h$.

Observe that for block sizes larger than the number of elements in the binary tree, \acmr[\csize][\len]
reduces to 
$\frac{\len}{\csize}$, a linear function of the edge length \len.
\comment{
\begin{equation}
  \label{eq:weightedsum}
  \acmr[\csize][\len] = \frac{\len}{\csize} \qquad (\len \leq \csize)
\end{equation}
}
This implies that the probability of a cache miss $\PB(\csize)$ reduces to 
$\frac{1}{W \csize} \sum_{ij \in E} \weight[ij] \len[ij]$, a weighted sum of the edge lengths.
\comment{
\begin{equation}
  \PB(\csize) = \frac{1}{W \csize} \sum_{ij \in E} \weight[ij] \len[ij]
              = \frac{\pwmean[1]}{\csize}
  \qquad (\len \leq \csize)
  \label{eqn:WLA}
\end{equation}
}
Thus, for very large block sizes, the optimal ordering is one that minimizes a weighted linear sum of edge lengths, where the weights are approximately geometrically decreasing as a function of the level of the edge.

\begin{figure*}[tp]
\centering%
\includegraphics[height=9pt]{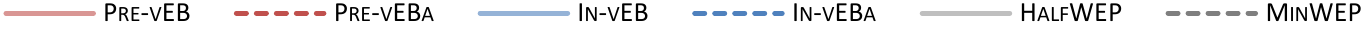}\\[1ex]%
\includegraphics[width=\columnwidth]{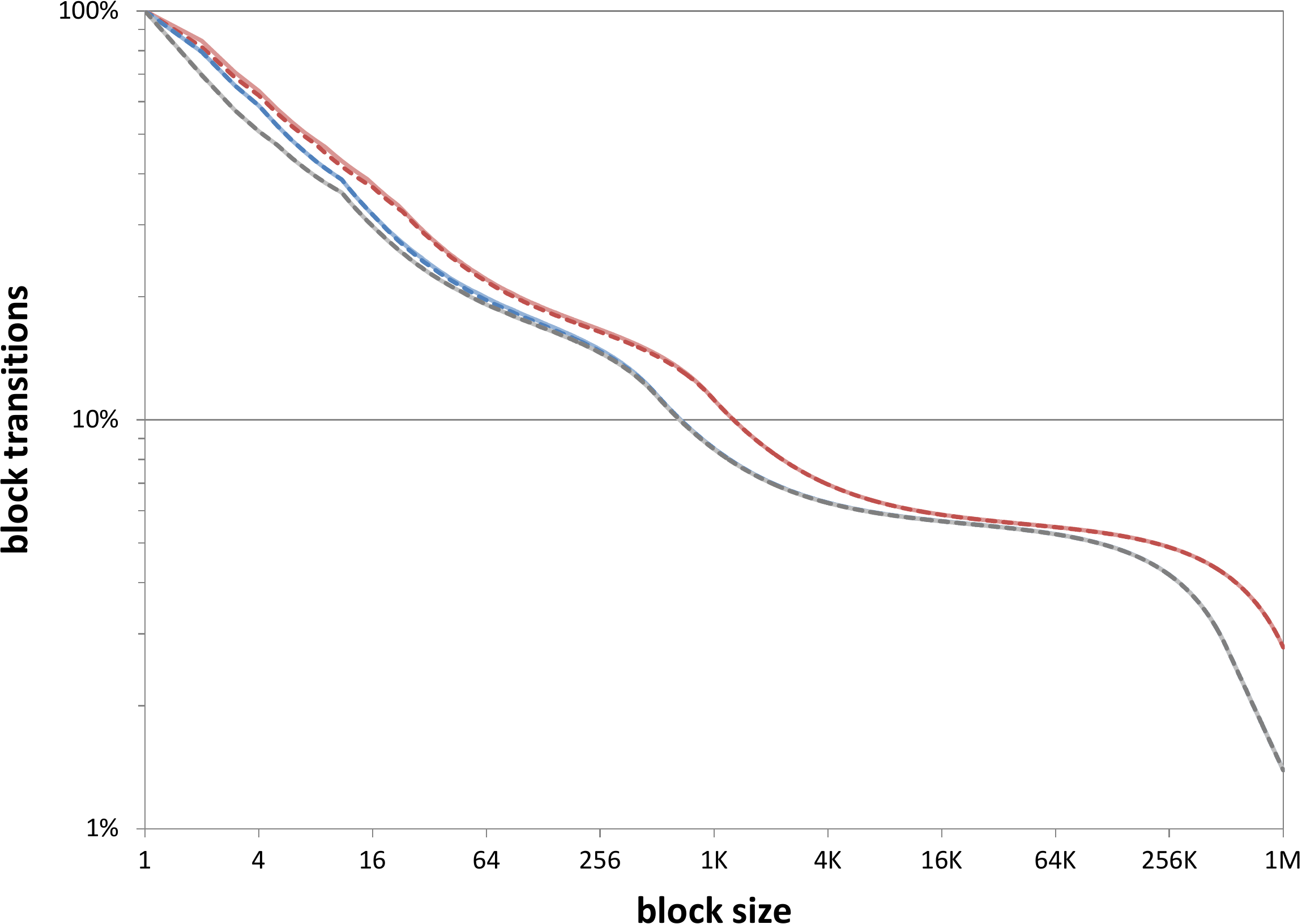}%
\hfill%
\includegraphics[width=\columnwidth]{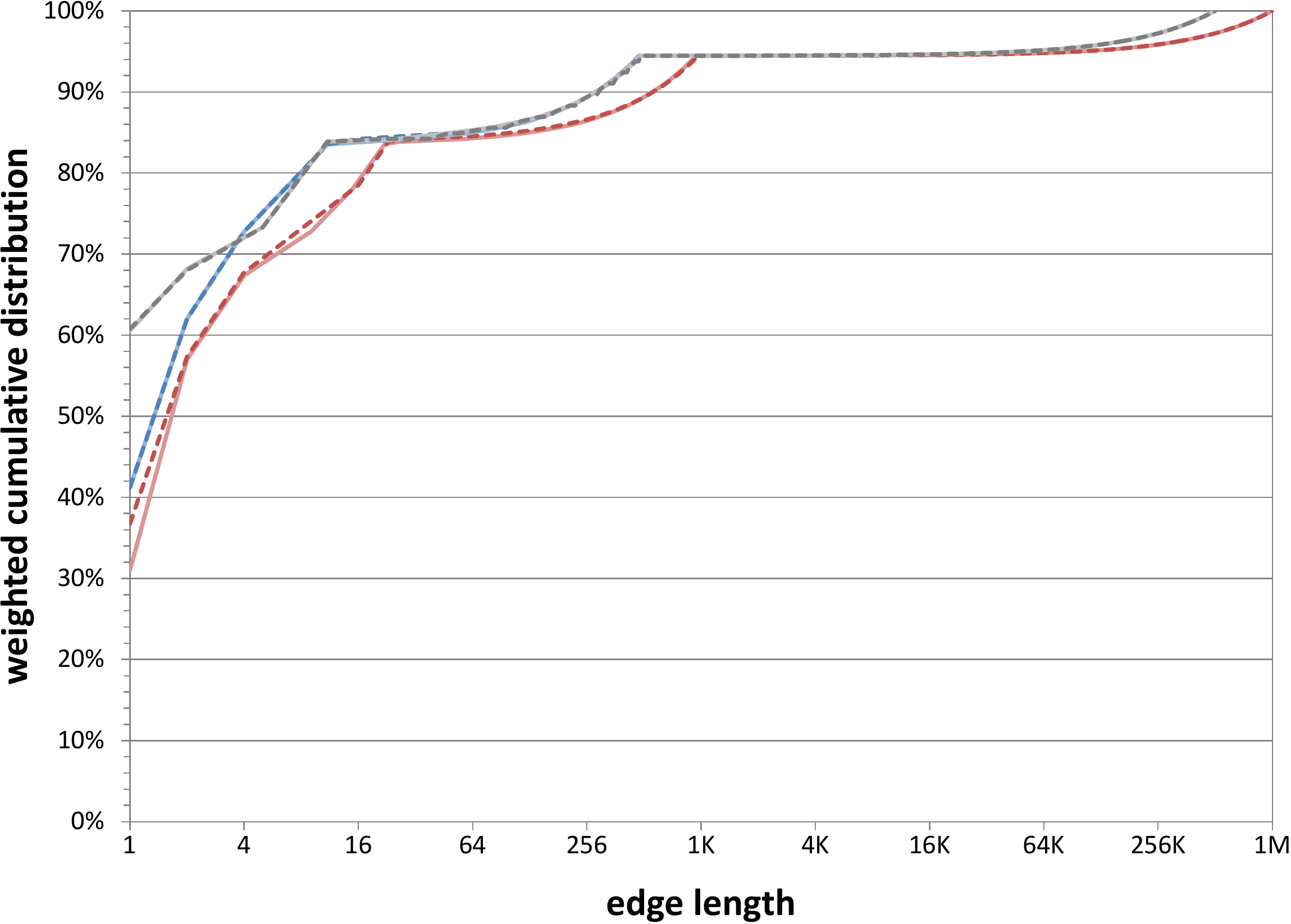}%
\caption{
  Two locality measures for several layouts of a tree of height $h = 20$.
  Left: Block transitions \PB as a function block size (lower is better).
  Right: Cumulative distribution of edge weights as a function of edge length (higher is better).
}
\label{fig:bt-in-pre}
\end{figure*}

\subsection{In-order \HLs}

Consider the in-order \vEBl, denoted as \ivl, and obtained by
arranging all bottom subtrees in-order, and in the same relative order as that of their parent leaves $L_A$. 
\comment{
wherein the top subtree lies in the middle of all the bottom subtrees.
following ordering of the subtrees, ordered consecutively in memory as $B_1, \ldots, B_{2^{g-1}}, A, B_{2^{g-1}+1}, \ldots, B_{2^g}$.
When $g = \FLOOR{h/2}$, we refer to this version as the in-order \vEBl, and 
which we denote as \ivl.
}
\comment{Observe that the bottom subtrees are not in boldface, since they are not going to be arranged in pre-order. We call the bottom subtrees $B_1, \ldots, B^{2^{g-1}}$ the left bottom subtrees, and others the right bottom subtrees}
In our nomenclature, \ivl is $\IN^{\FLOOR{h/2}}_1$ (see \autoref{tbl:notation}).
\comment{After the top subtree has been completely ordered, consider the leftmost $2^{g-2}$ parent leaves in $L_A$, and choose their children to be the left bottom subtrees. Once we choose the left and the right bottom subtrees, similar to \pvl, we arrange the bottom subtrees in \ivl in the same order as the order of their parent leaves $L_A$. (Later, in \autoref{sec:ivlpp}, we will see why this is the correct choice.) 
}
\autoref{fig:tree-vebi} illustrates \ivl for a tree of height $6$. Observe that at each branch of the recursion, the top subtree is arranged in-order. For instance, the \nodes on the top three levels are ordered in the middle of the layout, from positions $29$ to $35$. 
To compare \ivl with the pre-order \vEBl that arranges all subtrees pre-order (\pvl), we consider the percentage of block transitions \PB.
\comment{
So far, we have considered two versions of the \vEBl, either all subtrees arranged in-order (\ivl), or all subtrees arranged pre-order (\pvl). 
}

\autoref{fig:bt-in-pre} plots \PB for \pvl and \ivl as a function of block size for a tree of height 20. We see that \ivl dominates \pvl for every block size. Interestingly, even at very large block sizes, \ivl is much better than \pvl, indicating that it reduces the weighted average edge length.  We have observed the same dominance for trees of other heights.
In fact, for large block sizes, \ivl compares well with \minwep, which we introduce later as the optimal cache-oblivious \RL for binary search trees. 
Looking at the weighted cumulative distribution, which measures the total weight of all edges up to a certain length, we see the same dominance. 
Again, we observe that \ivl is indistinguishable from \minwep for large edge lengths.

\begin{figure*}[tp]
\centering%
\includegraphics[height=9pt]{veb-legend}\\[1ex]%
\includegraphics[width=\columnwidth]{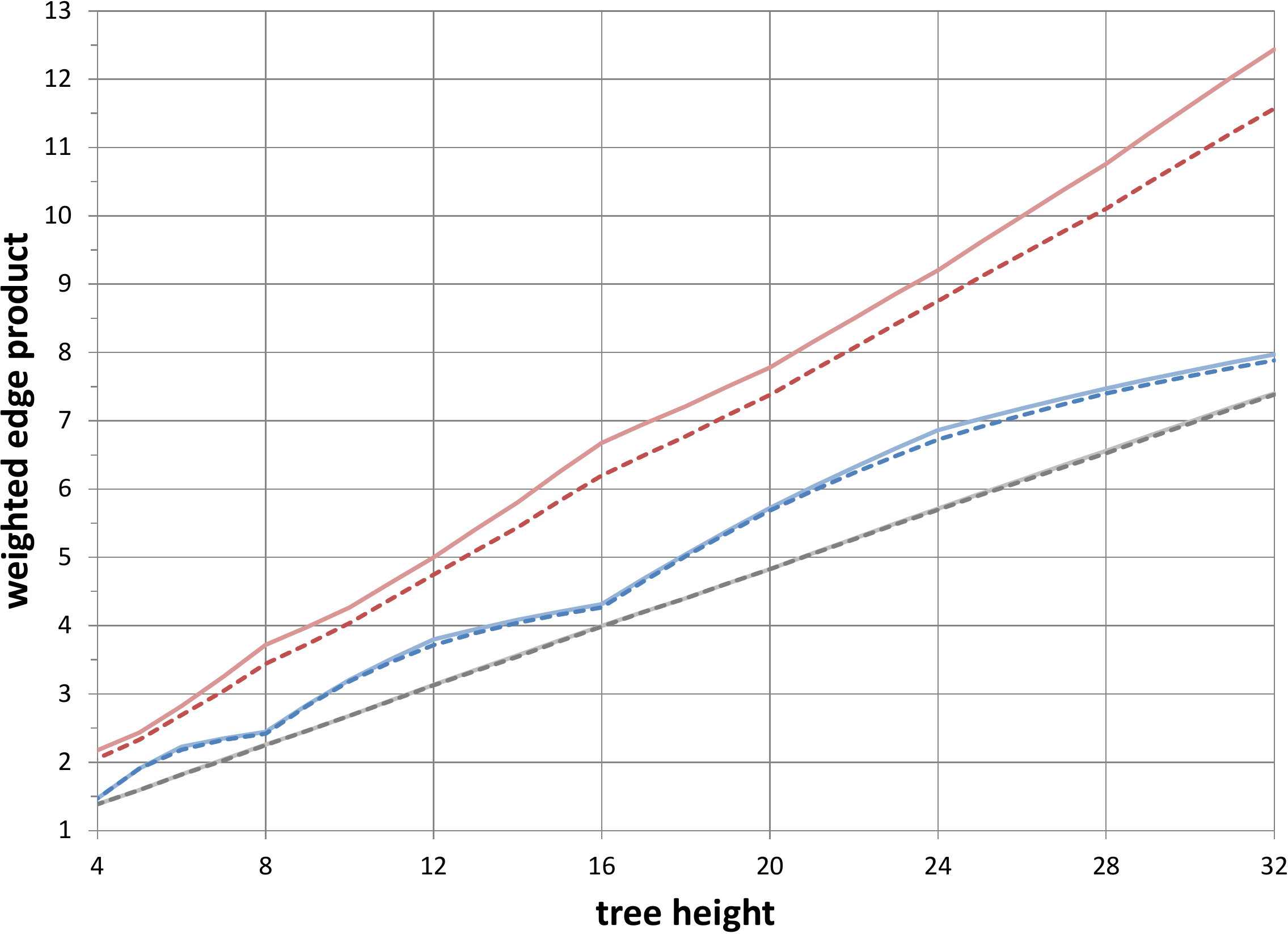}%
\hfill%
\includegraphics[width=\columnwidth]{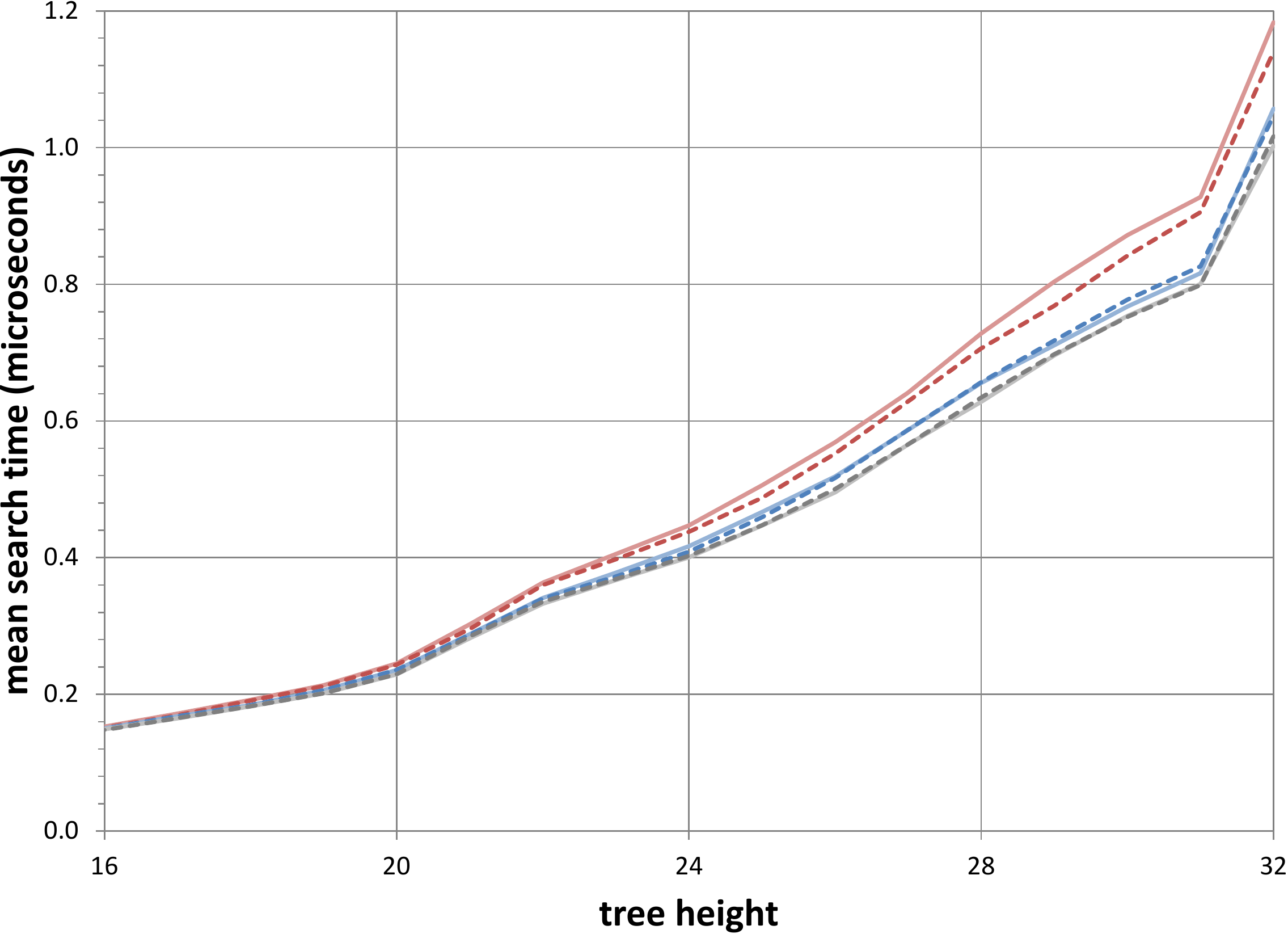}\\[2ex]%
\includegraphics[width=\columnwidth]{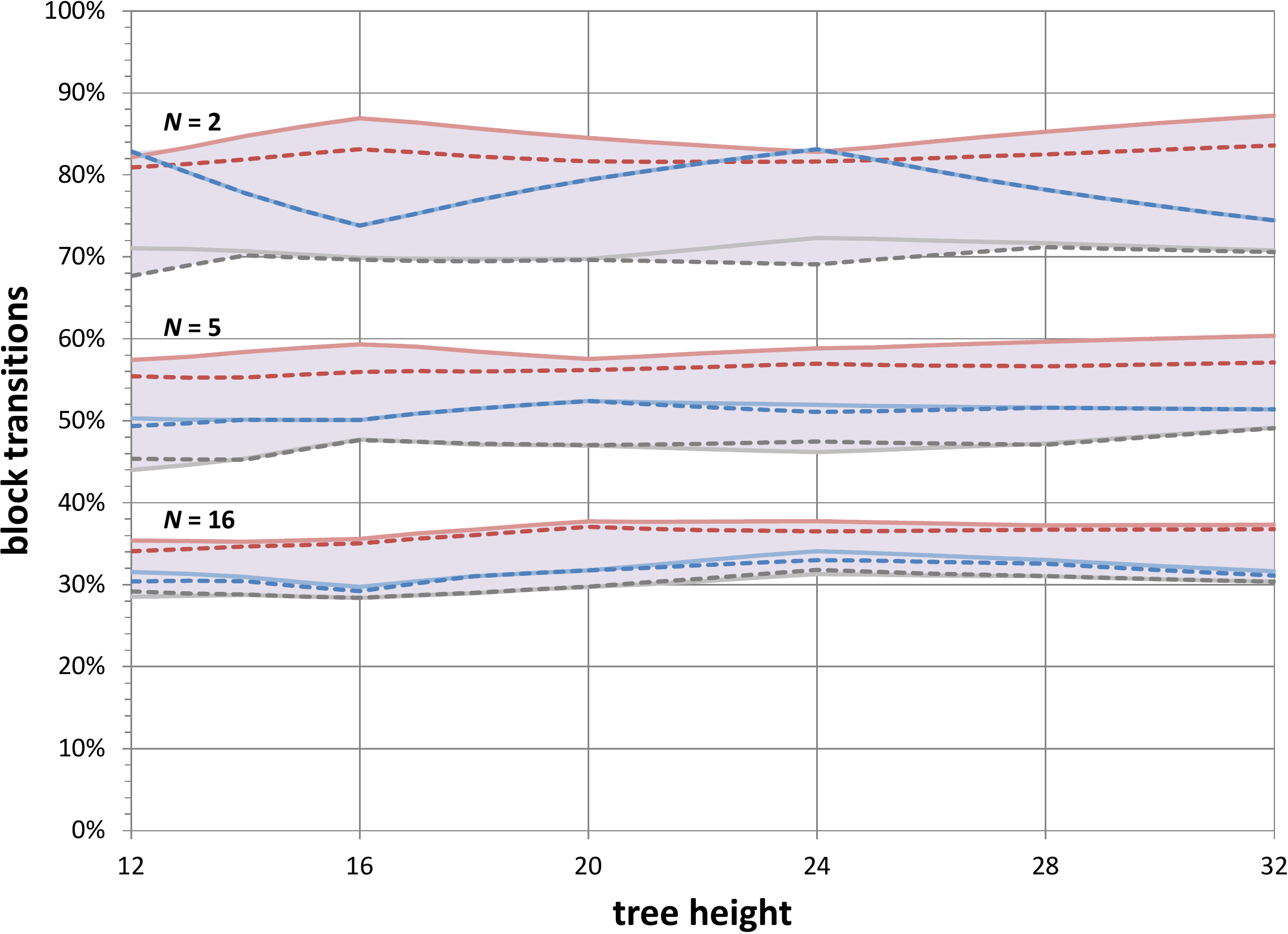}%
\hfill%
\includegraphics[width=\columnwidth]{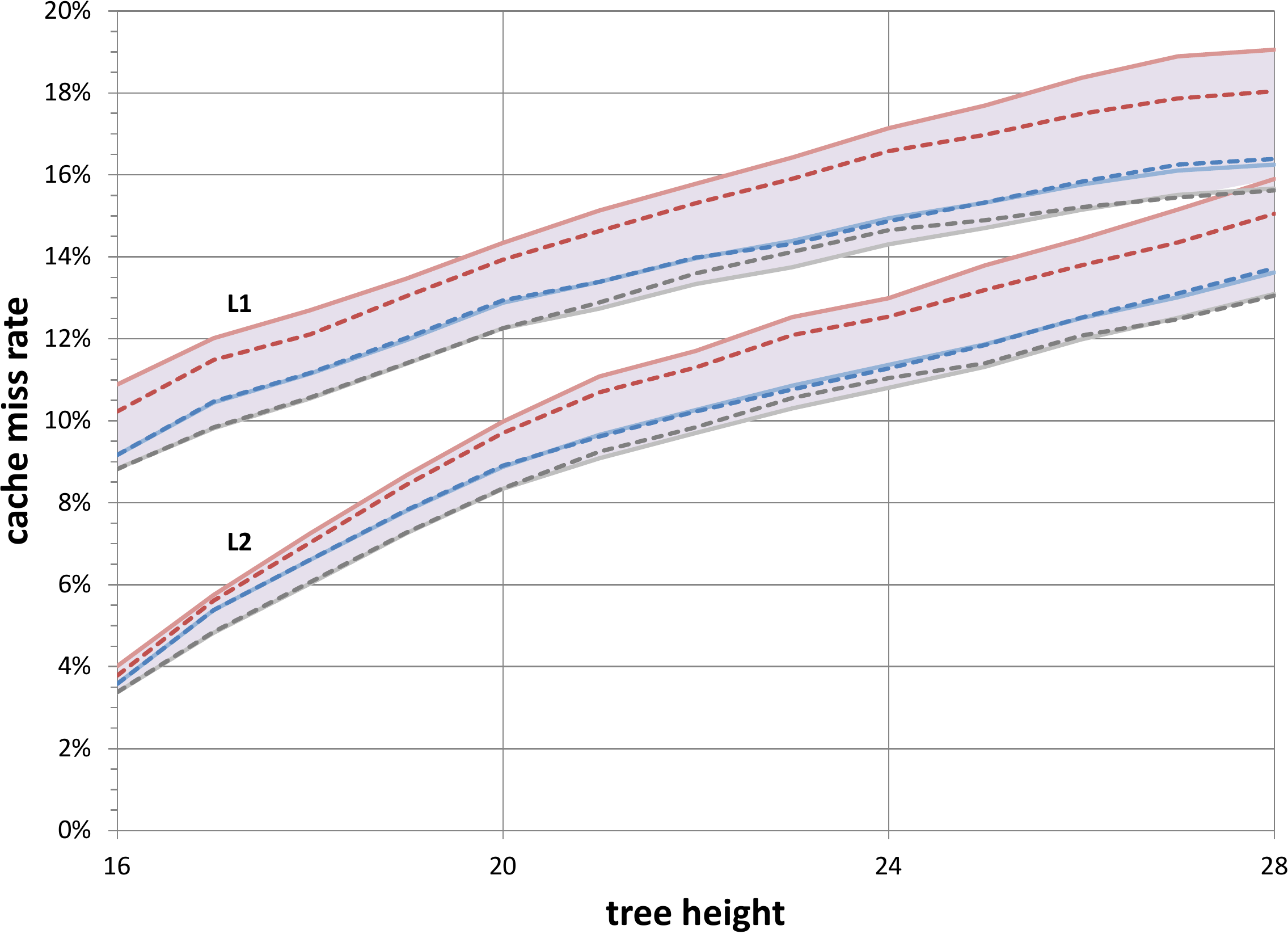}\\%
\caption{
  Clockwise from top left:
  weighted edge length product \pwmean[0];
  wall clock search time;
  L1 and L2 cache miss rate;
  and block transitions for blocks of $N \in \{2, 5, 16\}$ \nodes
  as a function of tree height for several hierarchical layouts.
}%
\label{fig:veb}%
\end{figure*}

\note{Point out that similar results were observed for different block sizes and tree heights. Stated.}
\autoref{fig:veb} plots \PB for \ivl and \pvl as a function of tree height for a block size of $2$, $5$, and $16$ \nodes. With $4$-byte \nodes, a block size of $16$ \nodes mimics a cache line size of 64 bytes. We see that \ivl dominates \pvl for all tree heights, but is dominated by \minwep. 
In our experiments, we observed similar results for other block sizes. 
\autoref{fig:veb} also illustrates the L1 and L2 cache miss rates for \ivl and \pvl. We observe the same dominance, and also that \minwep performs better than \ivl. Interestingly, \minwep results in even fewer L1 cache misses than the number of L2 cache misses for \pvl, suggesting that \minwep is a significantly better cache-oblivious layout than \pvl, the suggested layout in the literature. 

\note{Point out linearity of \pwmean[0] for \minwep? Not sure how to work this in.}
The true measure of any of these layouts is the average time taken to find any \node in the search tree (see \autoref{sec:setup} for more details on the experimental setup).
\comment{To measure this, we repeat the following experiment $10$ million times for each layout: Choose a \node at random, and measure the time taken to search for it.
}
To ensure that the wall clock search time is not affected by the time taken to compute the position of a \node in the layout, we store two child ``pointers'' with each \node. For this reason, we also refer to the search time as explicit, or pointer-based, search time. Illustrated in \autoref{fig:veb}, we see the same behavior as before. \ivl is significantly better than \pvl, but is marginally worse than \minwep. On average, \minwep is about $5\%$ better than \ivl and almost $20\%$ better than \pvl. The sudden uptick at $h=32$ is due to NUMA misses. Our experiments were run on a machine with two memory banks of $48$~GB each, and we need $64$~GB of RAM to store a tree of height $h=32$, generating a lot of traffic across the NUMA memory banks.  The plots in \autoref{fig:veb} therefore indicate that the percentage of block transitions (\PB) correlates very well with cache-miss ratios, and is therefore a good indicator of the quality of a layout. In \autoref{sec:wep}, we mathematically derive a new locality measure, the \WEP \pwmean[0], which is independent of the block size \csize and correlates even better with these measures and performance metrics. In \autoref{fig:veb}, we see that \ivl has much lower \pwmean[0] values than \pvl, but not as low as \minwep.

\comment{
Before we prove this for certain tree heights, we need to define the notion of a {\it level of detail}, as in \cite{Bender05}. Each level of detail is a partition of the tree into disjoint subtrees. In the finest level of detail, $0$, each \node forms its own subtree. In the coarsest level of detail, \CEIL{\log_2 h}, the entire tree forms the unique subtree.
\note{Only if $h$ is a power of two? Not really, works for any $h$.}
Level of detail $k$ is derived by starting with the entire tree, recursively partitioning it, exiting a branch of the recursion upon reaching a subtree of height less than or equal to $2^k$.
The key property of any \HL is that, at any level of detail, each subtree is stored in a contiguous block of memory. 
Observe that when the height of the tree is a power of $2$, and all subtrees are partitioned exactly in the middle ($g = h/2$), then all subtrees at level of detail $k$ are exactly of height $2^k$. 
\note{Above LOD $k=0$ was defined to be a tree of height 1.  Do you mean height $2^k$?. Yes, fixed.}
For every level of detail $k$, consider all the edges that connect the disjoint subtrees. Let this set be $E_k$. This set is the same for any \HL with cut height $g=h/2$. For a particular layout, let the \emph{tree distance} (measured in terms of number of disjoint subtrees spanned) of edge $e \in E_k$ be $\treelen{k}{e}$.  In other words, we treat each contiguous subtree of height $2^k$ as a single \node in a coarser tree of height $2^{-k} h$.

The weighted cumulative distribution in \autoref{fig:veb}, suggests that \ivl strongly dominates \pvl. 
To substantiate this, we state the following result. 
We delegate this and all subsequent proofs to 
the Appendix.
\begin{theorem}
\label{thm:in-pre-edgelen}
Consider a tree of height $h$, where $h$ is a power of $2$, and a level of detail $k$. For any edge $e \in E_k$, the tree distance $\treelen{k}{e}$ (measured as described above) for \ivl is no larger than the tree distance for \pvl.
\end{theorem}

\subsection{Properties of the in-order \vEBl}
\label{sec:ivlpp}

Given that \ivl is strictly better than \pvl, we study \ivl in detail, and ask whether \ivl can be improved even further.

We prove our results within the context of \HLs, which is essentially a framework of hierarchical 
decomposition into disjoint contiguous subtrees. 
One way to prove that one layout is better than another is to show that all tree distances decrease or remain the same for a given level of detail. First, we prove the choice of which bottom subtrees to place on either side of the top subtree $A$ in an \ivl layout. 
}

\comment{
Consider two leaves $x$ and $y$ of the top subtree. Let $X$ and $Y$ be bottom subtrees whose parents are
$x$ and $y$, respectively.  We will use $x < y$ to mean
$\pos[x] < \pos[y]$; $x < Y$ implies
$x < y \quad \forall y \in Y$; and $X < Y$ implies
$x < y \quad \forall x \in X, y \in Y$.  Without loss of generality,
assume $x < y$.

Consider the case in which two bottom subtrees $X$ and $Y$ appear
on either sides of the top subtree $A$ containing $x$ and $y$.
\begin{theorem}
Let $x$ and $y$ be leaves in the top subtree $A$ with 
child subtrees $X$ and $Y$, respectively.  If $x < y$, then for
any ordering with $Y < A < X$ there exists an ordering $X < A < Y$
with smaller tree distances.
\label{thm:subtree-in-order}
\end{theorem}

\begin{theorem}
Given any in-order subtree in a \HL and an internal ordering of the leaves of its top subtree $A$, the sum of the tree distances for all edges at a particular level of detail is reduced by arranging all the children of the left half of the parent leaves as the left bottom subtrees, and the rest as the right bottom subtrees.
\label{thm:subtree-in-order}
\end{theorem}
}

\comment{
As a corollary of \autoref{thm:subtree-in-order}, we see that the optimal division of the bottom subtrees into the left and right bottom subtrees is completely determined by the order of the leaves $L_A$ of the top subtree.
$A$ has an odd number of leaves only when $g=1$, \ie when $A$ is a single \node. In this case, the two bottom subtrees are ordered on either side of $A$. 
We have not yet determined what the optimal ordering of the bottom subtrees is for each of these sets (left and right) -- we have prescribed it to be in the order of the leaves. In \autoref{sec:alternate}, we return to this issue, and present an ordering that improves \ivl.
\sidenote{Try to get next section to start at top of page.}
\clearpage
}


\section{A Cache-Oblivious Locality Measure}
\label{sec:wep}

\todo{
Mathematical justification for edge-product. Simplify the general writeup to trees.
Define metric, justify metric. 
Review literature on other ``linear ordering metrics''. 

Memory transfer bounds: 
Do they depend on layout?
Do they depend on cut height?

Questions: Should all BSTs be inorder? Further we go from TST, inorder seems better. 
Restricting to in/pre.
Before we study further, try to understand the edge product measure.
}

\input{measure}

\input{weighted}

\section{Minimizing the Weighted Edge Product}
\label{sec:minwep}

\todo{

Experical claims:
First should be pre-order.
Second may be neighter, for lowest cost. Choose In for all the rest.

Present our grammar, that generalizes vEB layouts.

Experiments: block transitions, WEP cost.
Improvement in MinWEP.
Improvement in Block transitions.

Other interesting properties - Mean Edge distance.

Implementation issues: Implicit/explicit calculations.

Is this optimal? No. Neither in/pre may be the best. But we restrict to them. 
}

We have shown that layouts wth lower \WEP \pwmean[0] result in fewer block transitions (measured by \PB). So far, we have presented \ivl, with lower \pwmean[0] values than \pvl. 
\comment{
In the example presented in \autoref{fig:tree-layouts}, we see that \pwmean[0] for \ivl ($2.227$) is smaller than that of \pvl ($2.824$). 
}
\comment{A natural question to ask is: Is \ivl the best possible layout? }
\autoref{sec:alternate} shows that \pwmean[0] can be further reduced by \emph{alternating layouts}. 
\comment{In the rest of this paper, we introduce new \HLs that result in even lower \WEP.}
Ultimately, the goal is to find the \HL that minimizes \pwmean[0] -- the \minwep layout. 

\subsection{Ordering the subtrees: Alternating \HLs}
\label{sec:alternate}

\note{Should this not go in the section on minimizing \WEP?. Yes, moved there.}

In the discussion so far, we have not yet determined the optimal relative ordering of the bottom subtrees -- we have prescribed it to be in the order of the top subtree leaves. A simple way to reduce \pwmean[0] is to reduce the product of edge lengths among all edges that have 
the same weight, without modifying the lengths of all other edges.
\note{But alternating does change the sum of lengths, only the product. So what do we mean by reducing edge lengths? Fixed.}
If we consider the \HL at a particular branch of the recursion, all the edges between the top subtree and the bottom subtrees have the same weight.
By considering such equal-weight edges, the next result proves that a layout that orders the bottom subtrees in the reverse order of the parent leaves reduces \pwmean[0]. In such a layout,
the order of the \nodes appears to alternate between left-to-right and right-to-left. As a result, we refer to \HLs that arrange the bottom subtrees in the reverse order of the parent leaves as \textit{alternating} \HLs.
\comment{
Observe that one must cut and order the top subtree completely before ordering the bottom subtrees at each branch of the recursion. 
}
\comment{
As a consequence, \autoref{thm:in-pre-edgelen} proves that \ivl has a lower weighted edge product than \pvl, which manifests itself in lower \PB and better cache-oblivious layouts. 
}

\begin{theorem}
For any subtree in a particular branch of the recursion, suppose we fix the internal ordering of the leaves of the top subtree $A$ and the arrangement of all the bottom subtrees in subsequent branches of the recursion. Then, the product of all the edge lengths between the top subtree and the bottom subtrees is minimized by ordering the bottom subtrees in reverse order of that of the parent leaves $L_A$.
\label{thm:subtree-pre-order}
\end{theorem}

As a corollary of \autoref{thm:subtree-pre-order}, we see that when the cut height $g>1$, the optimal relative ordering of bottom subtrees is one that positions both the bottom subtrees of a particular parent leaf in $L_A$ adjacent to each other.
\note{This is not true when $g = 1$ and the root is in-order. Yes, fixed.}
This suggests that the initial orderings (\pvl and \ivl) got the adjacency of the bottom subtrees right -- they only had the order wrong.

By recursive application of \autoref{thm:subtree-pre-order}, we order the bottom subtrees in the reverse ordering of the parent leaves $L_A$ at each level of recursion, converting any \HL to its alternating version, thus reducing \pwmean[0].
\comment{
When we do this, the leaves of the tree appear to be alternating in order -- left-to-right at one level and right-to-left at the next -- explaining why we call such layouts alternating. Observe that this technique can be applied to any \HL, and the alternating version of any \HL reduces the \WEP. We denote the alternating version of any \HL by appending the letter \textsc{a}. In particular, we can define the alternating version of \ivl as the layout that arranges the bottom subtrees at each branch of the recursion in the reverse order of the parent leaves $L_A$ ---
}
We denote the alternating version of \ivl by \ivla, and define \pvla similarly. In our nomenclature, these two layouts are $\alt{\IN}^{\FLOOR{h/2}}_1$ and $\alt{\PR}^{\FLOOR{h/2}}_\infty$, respectively (see \autoref{tbl:notation}).

\autoref{fig:tree-vebia} illustrates \ivla for a tree of height $6$. 
Observe that the bottom subtrees are arranged in reverse order of their parent leaves.
In the outermost branch of the recursion, the rightmost two leaves in the top subtree are arranged at positions $35$ and $33$, and the corresponding child subtrees are rooted at positions $39$, $46$, $53$, and $60$. 
That is, the child subtrees are arranged in reverse order of the parent leaves ($39$ and $46$ connected to $35$), compared to 
\ivl (see \autoref{fig:tree-vebi}, where $39$ and $46$ are connected to $33$).
We see that by alternating, the sum of edge lengths between top and bottom subtrees remains the same, but we have increased their variance, thus reducing their product and consequently \pwmean[0].
A similar argument holds for alternating pre-order trees (see \autoref{fig:tree-veboa} and \autoref{fig:tree-vebo}).

\comment{
\autoref{fig:tree-veboa} illustrates \pvla for a tree of height $6$.
Observe that the bottom subtrees are arranged in reverse order of their parent leaves.
In the outermost branch of the recursion, the four leaves in the top subtree are arranged at positions $3$, $4$, $6$, and $7$, and the corresponding child subtrees are rooted at positions $8, 15, 22, \ldots, 57$. Compare this with \pvl (see \autoref{fig:tree-vebo}) where these child subtrees are arranged in the same order as the parent leaves.
As in the previous comparison, we notice that the \pwmean[1], which measures the weighted sum of edge lengths, has not changed. However, the \WEP has been reduced. 
\note{The number of unit-length edges *does* increase in \pvla. Added.}
}

It is important to mention that \textit{alternating} a particular layout has no effect on \pwmean[1]. However, since the variance of the edge lengths is increased, alternating a layout will increase \pmean[\infty], and may increase the number of unit-length edges. Since \pwmean[0] is a function of the product of the edge lengths, the \WEP is reduced. As an example, we can see the effect of alternating a layout on \pwmean[0], \pwmean[1], and \pmean[\infty]
by comparing \ivla (\autoref{fig:tree-vebia}) and \ivl (\autoref{fig:tree-vebi}). 
\note{See what? Fixed.}
 
\autoref{fig:veb} shows that \pvla has smaller \pwmean[0] values than \pvl, but this improvement is not as drastic as the improvement from \pvl to \ivl. A natural question to ask is: Does an ordering that reduces \pwmean[0] result in better cache-oblivious layouts? And if so, do we get a greater improvement from \pvl to \ivl, as predicted by the \pwmean[0] values? We first look to block transition percentages (\PB) to show that alternating layouts are better.
\autoref{fig:bt-in-pre} plots \PB for \pvla and \ivla as a function of block size for a tree of height 20. We see that \ivla is virtually indistinguishable from \ivl, whereas \pvla dominates \pvl for small block sizes. 
\autoref{fig:veb} also plots \PB for \pvla and \ivla as a function of tree height for a variety of block sizes. 
\comment{for these block sizes, }
Again, 
we see that \ivla is virtually indistinguishable from \ivl, but \pvla dominates \pvl for all tree heights. And we see the same pattern with cache miss rate. \autoref{fig:veb} also plots the explicit search time for \ivla and \pvla. Comparing with \ivl and \pvl, we see the exact same pattern. \ivl and \ivla are indistinguishable from each other, with approximately $5\%$ worse explicit search times than \minwep. On the other hand, \pvla is about $5\%$ better than \pvl. 

From these experiments, we see that the improvement from \pvl to \pvla is far less than the improvement from \pvl to \ivl. This suggests that while an alternating version always improves the layout
(we restrict our attention to alternating layouts for the rest of this paper), it is far more important to switch from pre-order to in-order. One should consider this the main take-home message of this paper: All data structures that use a pre-order \HL should, at the very least, switch to an in-order version of the same \HL. Later, in \autoref{sec:cut-hts}, we will see that this result may depend on the cut height, but not for the cut heights $\FLOOR{h/2}$ that have been used in practice. 

\subsection{Constructing hybrid layouts: The \halfwep layout} 

\comment{A more critical question remains: What should be the position of}
We now analyze the impact of varying the position of the top subtree $A$ relative to all the bottom subtrees.
\RLs restrict us to the two extremes represented by \pvla and \ivla, wherein $A$ is positioned either at one end or in the middle of all the bottom subtrees. However, \ivla and \pvla arrange \textbf{all} bottom subtrees identically, 
either in-order or pre-order, respectively. 
\note{We need to make it clear that pre-order means both pre- and post-order depending on context. Added text in the definition pf \HLs.}
We can consider many more permutations by ordering some of the bottom subtrees in-order and others pre-order. As before, 
the locality measure \pwmean[0] guides us in these decisions. Clearly, \ivla results in smaller \pwmean[0] than \pvla. 
Can a hybrid layout (by modifying \ivla, possibly) reduce \pwmean[0] even further? 

To construct a hybrid layout, we must take into account the trade-offs involved. First, observe that any bottom subtree is arranged in a contiguous block in memory, and has only one edge connecting it to the rest of the tree -- the edge from its root to a leaf in the top subtree. Therefore, rearranging any bottom subtree potentially results in two changes to its contribution to \pwmean[0]: the length of the edge connecting its root to its parent, and the lengths of the edges in the subtree itself. Discounting the connection to the top subtree, a bottom subtree ordered as in \pvla has a larger \WEP than when it is ordered as in \ivla. However, the root of a pre-order bottom subtree is closer to its parent than the root of an in-order bottom subtree, and the weight of this edge is larger than the weight of any edge within the bottom subtree. So there are potential benefits to modifying \ivla by arranging some of the bottom subtrees pre-order. This nevertheless raises the question: Which bottom subtrees should we modify, if any? Also, observe that the in-order bottom subtrees are identical, and differ only in their distance to their parent leaf in the top subtree, and similarly for the pre-order trees.
As we move further away from the top subtree, the proportional reduction in the length \len of the edge connecting the subtrees decreases (\ie the slope of $\log \len$ approaches zero), whereas the degradation in the \pwmean[0] value of the bottom subtree remains the same. As a result,
the marginal benefit of converting an in-order bottom subtree into a pre-order bottom subtree decreases. Therefore, if arranging any bottom subtree in-order results in lower \pwmean[0] than arranging it pre-order, then this must also be true for all bottom subtrees further away from the top subtree.

\note{Do we need to discuss dynamic trees (insertions/deletions) or incomplete trees? I don't think we need to, since we refer to Bender's papers that present dynamic search trees. And it is not the scope of this paper anyway.}
To find the best layout, we undertook a detailed empirical study that evaluated all \RLs for trees up to height $20$.
We considered all possible cut heights $g \leq \FLOOR{h/2}$ (we quickly determined that larger $g$ were not beneficial). We calculated \pwmean[0] for every
layout 
for each tree height. We noticed that the optimal ordering always
arranged the bottom subtrees closest to the top subtree pre-order, arranged all other bottom subtrees in-order, and used an in-order arrangement for the outermost branch of the recursion. In comparison, \ivla arranges all bottom subtrees in-order, and \pvla arranges all of them pre-order. We give the version of this layout with cut height $g = \FLOOR{h/2}$ the special name \halfwep. In our nomenclature, \halfwep is $\alt{\IN}^{\FLOOR{h/2}}_2$ (see \autoref{tbl:notation}).

\begin{figure*}[tp]
\centering%
\includegraphics[height=7.5pt]{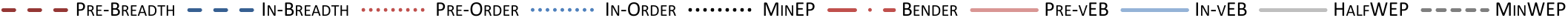}\\[1ex]%
\includegraphics[width=\columnwidth]{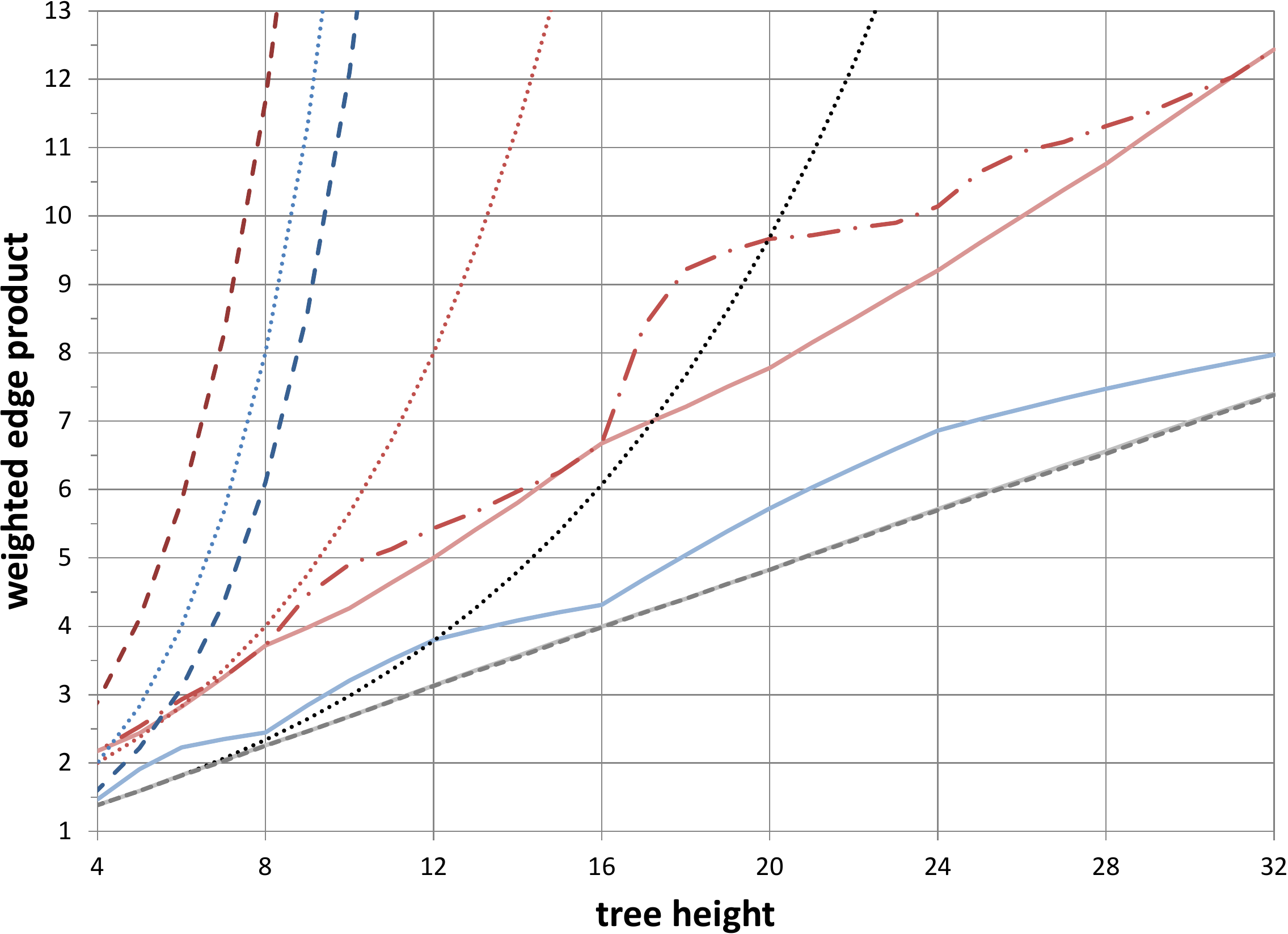}%
\hfill%
\includegraphics[width=\columnwidth]{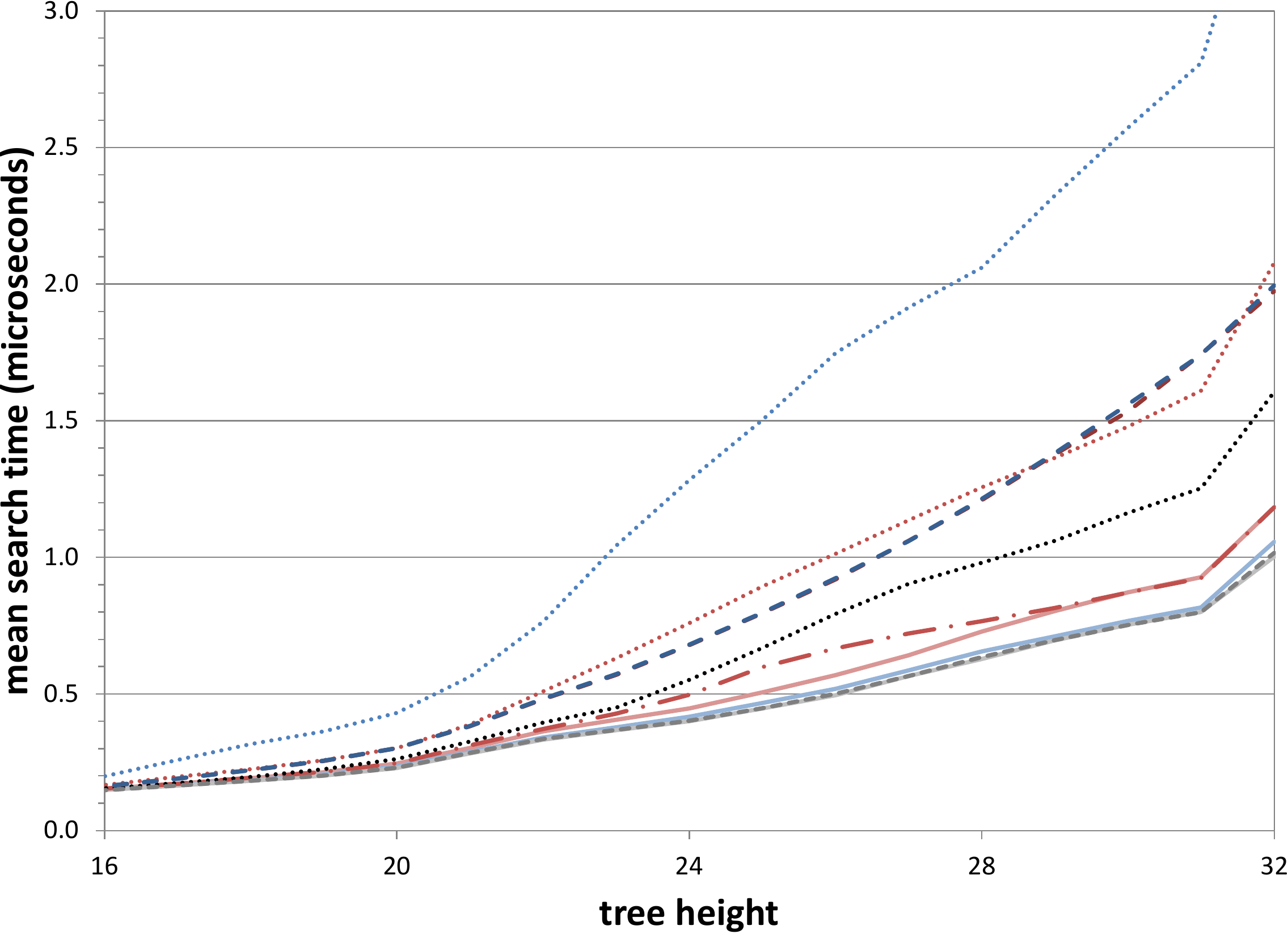}\\[2ex]%
\includegraphics[width=\columnwidth]{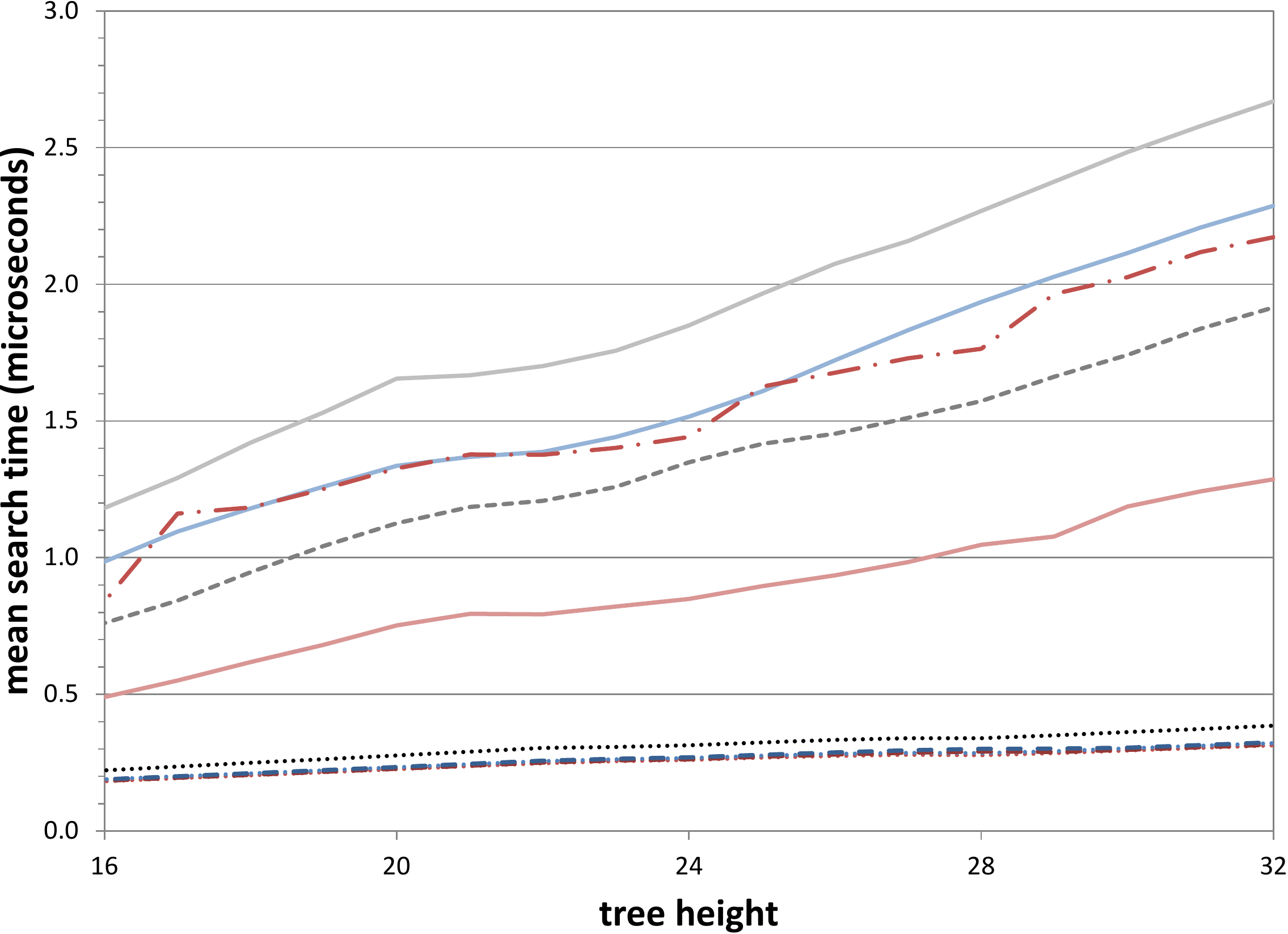}%
\hfill%
\includegraphics[width=\columnwidth]{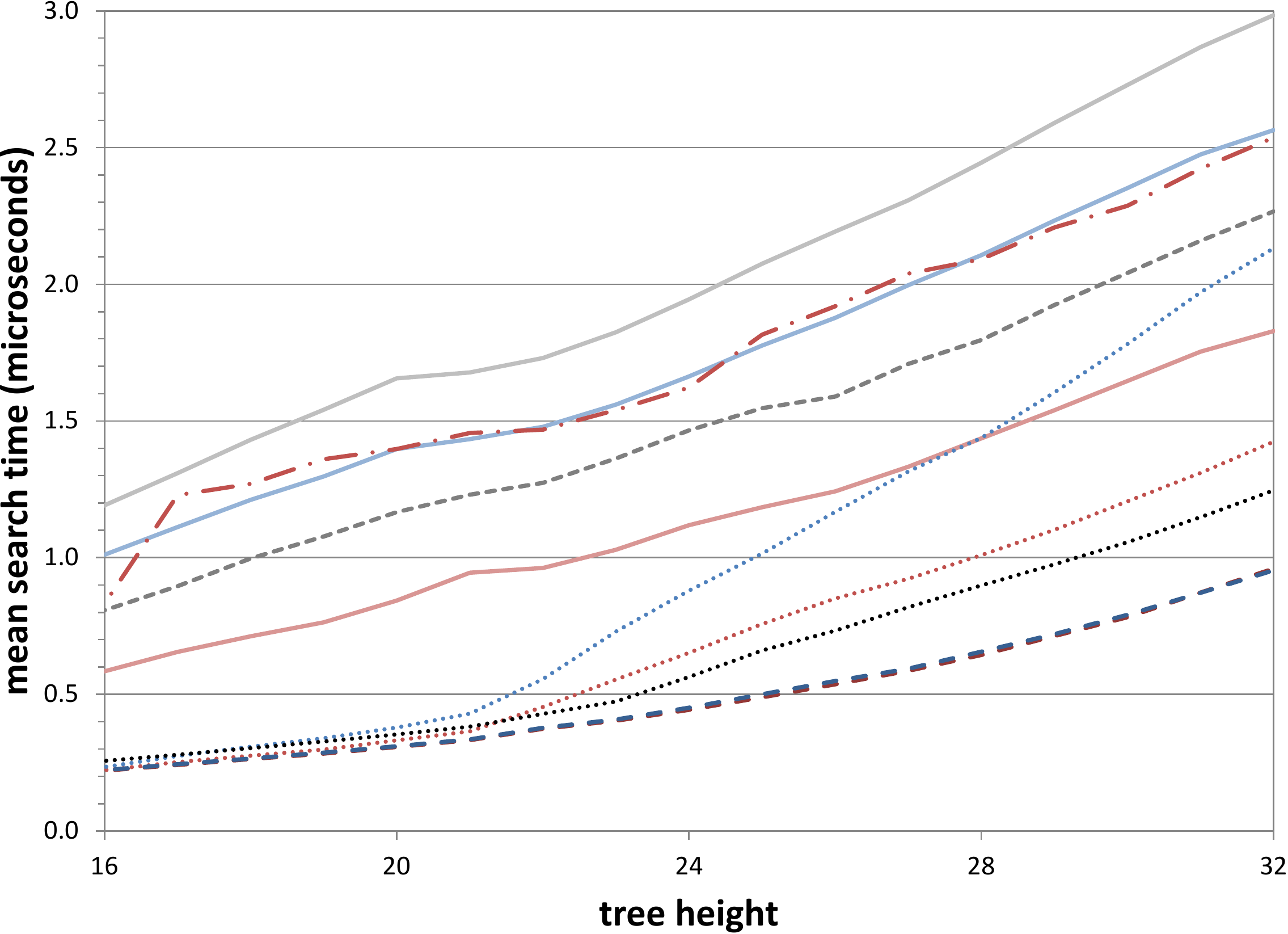}\\%
\caption{
  Clockwise from top left:
  weighted edge length product \pwmean[0];
  pointer-based search time;
  pointer-less search time;
  and pointer-less search time excluding all memory accesses
  as a function of tree height for several layouts.
}%
\label{fig:cut}%
\end{figure*}

\autoref{fig:cut} shows that \halfwep and \minwep have almost indistinguishable values of \pwmean[0] and performance in explicit search times, further validating \pwmean[0] as the appropriate locality measure for deriving cache-oblivious layouts. 

\autoref{fig:tree-halfwep} illustrates \halfwep for a tree of height $6$. Observe that the bottom subtrees closest to the top subtree are arranged pre-order. 
\comment{are the closest to the top subtree }
At the outermost branch of the recursion, the subtrees rooted at positions $28$ and $36$ 
are arranged pre-order in \halfwep. These are arranged in-order in \ivla (see \autoref{fig:tree-vebia}). From the thickness of the edges, one can see that \halfwep reduces some edge lengths for every branch of the recursion by replacing some in-order bottom subtrees by pre-order bottom subtrees. This does increase some distances within the bottom subtree (the next recursive branch), but deeper down the tree, where they contribute less to the \WEP. This is confirmed by the \pwmean[0] values for \halfwep ($1.823$) and \ivla ($2.184$).

\note{Need to talk about \halfwep and \minwep almost identical. Done.}

\comment{
We believe that if we restrict ourselves to \HLs with cut height $g=\FLOOR{h/2}$,
then \halfwep minimizes \pwmean[0]. This is a conjecture, based on some theoretical results and empirical observations. Empirically, based on experiments on trees up to a height of $32$, we observed that this is true.
}
\note{We did? No. Changed sentence accordingly.}
Our empirical analysis is also backed by theory, when restricted to certain cut heights. In \autoref{thm:order}, we show that when the cuts are made at the top of the tree ($g = 1$) at all branches of the recursion, this \halfwep-like layout provably minimizes \pwmean[0]. 
\comment{
In this layout, a pre-order subtree is ordered as
$A, \bm{B_1}, B_2$, and an in-order subtree is ordered as $\bm{B_1},  A, \bm{B_2}$.}
We refer to this layout as the \minep layout, because it also minimizes the edge product \pmean[0] for unweighted trees.
In our nomenclature, \minep is $\IN^1_2$ (see \autoref{tbl:notation}).

\begin{theorem}
\label{thm:order}
The \minep layout minimizes \pwmean[0] among all \RLs with cut height $g=1$.
\end{theorem}
 
It can also be shown that for all \HLs where the closest bottom subtree is arranged pre-order, cutting a subtree that is arranged in-order at $g=1$ results in the same layout as cutting it at $g=2$. This is because cutting an in-order subtree of height $h$ at height $g=1$ results in two pre-order subtrees of height $h-1$, the roots of which are adjacent to the top subtree, so long as the closest bottom subtree in either case is arranged pre-order. 
\comment{
This is equivalent to cutting the top subtree at height $g=2$, 
}
This explains why, for trees of small height, such as the example with $h=6$ considered in \autoref{fig:tree-layouts}, \minep is identical to \minwep (see \autoref{fig:tree-minwep}).

\subsection{Optimizing the cut height: The \minwep layout}
\label{sec:cut-hts}

In the discussion so far, we have ignored the effect of the cut height by restricting ourselves to the case where $g=\FLOOR{h/2}$. Before we find the optimal cut height, we describe other layouts that turn out to be part of the \HL framework, albeit with extreme cut height values.

Consider cut height $g=1$. Analogous to how \halfwep is a hybrid of \ivla and \pvla, one can think of \minep as a hybrid of two other simple layouts: the common \inorder and \preorder depth-first layouts. All three are \HLs that cut every subtree at height $g=1$, but differ in how the bottom subtrees are arranged. \inorder arranges all subtrees in-order, and \preorder arranges all subtrees pre-order. In our nomenclature, \inorder is $\IN^{1}_1$ and \preorder is $\PR^1_\infty$ (see \autoref{tbl:notation}).
Observe that when the cut height $g=1$, there is only one leaf \node in $L_A$ at every branch of the recursion, and therefore the notion of alternating layouts is not relevant. Furthermore, there are only $2$ bottom subtrees at each branch of the recursion, and therefore in-order and pre-order are the only two options for positioning the top subtree. 
\comment{Due to this restriction, }
As a result, all \HLs that are described using cut height $g=1$ at all branches of the recursion are in fact \RLs. This is not true for other cut heights.

\autoref{fig:tree-inorder} illustrates \inorder for a tree of height $6$. Observe that \inorder arranges the two bottom subtrees at the outermost recursion in-order, resulting in their roots being placed at positions $16$ and $48$. These roots are arranged pre-order at positions $31$ and $33$ in \minep (see \autoref{fig:tree-minep}). On the other hand, \preorder (see \autoref{fig:tree-preorder}) arranges all subtrees pre-order. The roots of the same bottom subtrees are arranged pre-order at positions $2$ and $33$. 
Observe that \inorder and \preorder have (nearly exactly) the same number of short edge lengths (counting the number of thick lines), but these are at the bottom of the tree for in-order, where the weights are much smaller. This results in much larger \pwmean[0] values for \inorder ($4.000$), when compared to \preorder ($2.828$).

At the other end of the spectrum in terms of cut height is $g=h-1$, where each subtree is cut one level above the bottom. 
It turns out that the \HL with $g=h-1$ that arranges all subtrees pre-order (similar to \pvl with cut height $g=\FLOOR{h/2}$) is the simple and commonly used breadth-first order. For this reason, we denote the breadth-first layout as \prebreadth. \autoref{fig:tree-breadth} illustrates the \prebreadth layout for a tree of height $6$. Observe that the \nodes are arranged by level. Furthermore, one can now also consider in-order and/or alternating variants on the breadth-first ordering. We denote the in-order variant by \inbreadth. Observe that when $g=h-1$, the bottom subtrees are single \nodes, and therefore the notion of their arrangement into pre- or in-order is not relevant. In our nomenclature, 
\inbreadth is $\IN^{h-1}_*$ and \prebreadth is $\PR^{h-1}_{*}$ (see \autoref{tbl:notation}).

\todo{
What about cut heights? Emperical graph comparing with optimal.

ApproxWEP: in/pre have different cut heights. Why do we use these? Motivation - symmetric, and empirical.
Other Observations: For MINWEP, special cases cut heights.

}

\comment{
To find the best cut height, we undertook a detailed empirical study that computed \pwmean[0] for all \HLs restricted to in-order and pre-order arrangements of the top subtree at each branch of the recursion. Two factors were allowed to vary. We allowed every subtree to be aranged in-, pre-, or post-order, and we considered all possible cut heights $g \leq \FLOOR{h/2}$ (we quickly determined that larger $g$ were not beneficial). We calculated \pwmean[0] for each potential layout to find \minwep for each tree height. 
}
In our detailed empirical analysis, which suggested that \pwmean[0] is minimized by layouts that fit the characterization $\alt{\IN}^*_2$ in our nomenclature, we noticed that the optimal cut height (denoted by \opt) was different from \halfwep for pre-order subtrees: $g_P^{\opt}(h) = \max\{1, \FLOOR{(h-1)/2}\}$.
\comment{
\begin{itemize}
\item
For every case, the \minwep layout arranged the bottom subtrees just as in \halfwep --- Those closest to the top subtree were arranged pre-order, and all others in-order.
\item
}
For in-order subtrees, it is the same as before, \ie, $g_I^\opt(h) = \FLOOR{h/2}$.
\comment{The optimal cut height, denoted by \opt, is different from \halfwep for pre-order subtrees}
Furthermore, there is one exception to the above rule, with $g_I^\opt(6) = 2$, and correspondingly
$g_P^\opt(5) = 1$.

\comment{
\begin{table}
\newcommand{\m}[1]{$\mathbf{#1}$}
\centering
\begin{tabular}{|r|c|c|c|c|}
\hline
\multicolumn{1}{|c|}{\multirow{2}*{$h$}} &
\multicolumn{2}{c|}{$\weight[i] = 2^{-i}$} &
\multicolumn{2}{c|}{$\weight[i,h] = \frac{\nodeset[h-i]}{\nodeset[h]}$} \\
\cline{2-5}
& 
\multicolumn{1}{c|}{in} &
\multicolumn{1}{c|}{pre} &
\multicolumn{1}{c|}{in} &
\multicolumn{1}{c|}{pre} \\
\hline
 2 &   $1|1$ & $1|1$ &  $1|1$ &  $1|1$ \\
 3 &   $1|2$ & $1|2$ &  $1|2$ &  $1|2$ \\
 4 &   $2|2$ & $1|3$ &  $2|2$ &  $1|3$ \\
 5 &   $2|3$ & \m{1|4} &  $2|3$ &  $2|3$ \\
 6 & \m{2|4} & $2|4$ &  $3|3$ &  $2|4$ \\
 7 &   $3|4$ & $3|4$ &  $3|4$ &  $3|4$ \\
 8 &   $4|4$ & $3|5$ &  $4|4$ &  $3|5$ \\
 9 &   $4|5$ & $4|5$ &  $4|5$ &  $4|5$ \\
10 &   $5|5$ & $4|6$ &  $5|5$ &  $4|6$ \\
11 &   $5|6$ & $5|6$ &  $5|6$ &  $5|6$ \\
12 &   $6|6$ & $5|7$ &  $6|6$ &\m{6|6} \\
13 &   $6|7$ & $6|7$ &\m{7|6} &  $6|7$ \\
14 &   $7|7$ & $6|8$ &  $7|7$ &  $6|8$ \\
15 &   $7|8$ & $7|8$ &  $7|8$ &  $7|8$ \\
16 &   $8|8$ & $7|9$ &  $8|8$ &\m{8|8} \\
\hline
\end{tabular}
\caption{
  Optimal partitions $t|b$ of a tree of height $h = t + b$ using
  geometric and exact weights for in- and pre-/post-order
  positioning of the top subtree.  $t$ denotes the height of the
  top subtree.  Exceptions to the simple partitioning rules
  ($t = \lfloor h/2 \rfloor$ for in-order;
  $t = \max\{1, \lfloor (h-1)/2 \rfloor\}$
  for pre- and post-order) are highlighted.
}
\label{tbl:partitions}
\end{table}
}
\note{Is this \minwep? Yes.}

Based on these experiments, we define \minwep as the \halfwep-like layout with the cut heights presented above, including the exception. In our nomenclature, \minwep is $\alt{\IN}^\opt_2$ (see \autoref{tbl:notation}). \autoref{fig:tree-minwep} illustrates \minwep for a tree of height $6$. 
In the outermost branch of the recursion, the top two levels of the tree are arranged together in positions $31$ to $33$, indicating a cut of $g=2$. This compares with a cut of height $g=3$ for \halfwep (see \autoref{fig:tree-halfwep}). Also, we see that the pre-order subtree of height $h=4$ rooted at position $34$ in \minwep is cut at a height $g=\FLOOR{(h-1)/2}=1$.
\comment{
We also denote the layout that does not include the exception as \approxwep. Thus \approxwep is a layout that arranges the bottom subtrees closest to the top subtree pre-order and cuts them at height $g_{apx,P}=\max\{1,\FLOOR{(h-1)/2}\}$. It arranges all other subtrees in-order and cuts them at height $g_{apx,I}=\FLOOR{h/2}$, just like \halfwep does. 
}

The pre-order cut height exception at $g_P^\opt(5)$ can also be interpreted as part of the piece-wise function 
$g_P^\opt(h) = 1$ if $h \leq 5$, and $\FLOOR{(h-1)/2}$ otherwise. Furthermore, interpreting the in-order cut height for subtrees of height $3$ as $g=2$ instead of $g=1$, which is an equally valid interpretation, the cut height for an in-order subtree can be calculated directly from the cut height for an pre-order subtree as follows: $g_I^\opt(h) = 1$ if $h=2$, and $g_P^\opt(h-1) + 1$ otherwise. 

Analyzing \HLs where the closest bottom subtree is arranged pre-order and the cut heights are chosen such that $g_I(h) = g_P(h-1) + 1$, we see that cutting all in-order subtrees at height $g_I(h)=1$ instead results in the same layout. This is because this in-order cut results in two pre-order bottom subtrees of height $h-1$, each of which will subsequently be cut at the same height as they would have been if they had been part of an in-order subtree of height $h$. Since the closest bottom subtree is pre-order in either case, the layouts are identical. As a result, we can set $g_I^\opt(h) = 1$. As we shall see later, this is important since it simplifies the index computation for \nodes in pointer-less trees. 
Note that this optimization cannot be applied to \halfwep.
\comment{, since its pre-order and in-order cut heights are identical.}

\comment{
For small tree heights upto $h=6$, \approxwep is identical to \halfwep, and is not separately illustrated in \autoref{fig:tree-layouts}. For larger trees, \approxwep differs from \halfwep and \minwep slightly, and tends to perform similar to both.We have already seen from \autoref{fig:bt-in-pre} and \autoref{fig:veb} that in all the metrics considered, \halfwep and \approxwep already perform more or less identically. For this reason, we do not include \approxwep in our figures, even though it is different from \minwep.
}

\subsection{The cost of cache-misses: Explicit pointer-based searches}

We observed earlier that \halfwep and \minwep are virtually indistinguishable in terms of explicit search time. This is because they are exactly the same ordering schemes, but with very slightly different cut heights. Larger differences in cut heights can make a significant difference. Consider the values of \pwmean[0] in \autoref{fig:cut} for many of the layouts presented so far. Recall that \bender and \pvl differ from each other only in the choice of the cut height $g$. For \bender, the cut height $g = h - 2^{\CEIL{\log_2(h/2)}}$, which is identical to \pvl ($g = \FLOOR{h/2}$) only for subtree heights that are a power of two. As expected, we see identical values of \pwmean[0] for \bender and \pvl for trees of height $4$, $8$, $16$, and $32$. However, for all other tree heights, \bender gives higher values for \pwmean[0]; sometimes $20\%$ worse. This manifests itself in similarly worse pointer-based search times compared to \pvl. This suggests that for a particular ordering scheme, the optimal cut height is closer to halfway down the tree.
\comment{
for all ordering schemes (combination of in- and pre-order subtrees in the same column in \autoref{tbl:notation}).
}

Cut heights $g=1$ and $g=h-1$ illustrate this further. \minep, which is identical to \minwep except in its choice of the cut height ($g=1$), results in significantly different trees (especially for larger tree heights), and we observe a steep divergence in \pwmean[0] as the tree height increases. 
As expected, \minep's performance (measured using pointer-based search times) also degrades significantly for large tree heights. 
At the other end of the spectrum, consider \prebreadth and \inbreadth, which are identical to \pvl and \ivl respectively, except in the choice of cut height. A cut height of $g=h-1$ results in significantly different layouts, especially for large tree heights. Even for the small example in \autoref{fig:tree-layouts}, we see that \prebreadth is quite different from \pvl. In \autoref{fig:cut}, we see that the pointer-based search time is significantly worse for breadth-first layouts, when compared to \pvl and \ivl. 

From the \pwmean[0] values in \autoref{fig:cut}, we also observe that in-order is not always better than pre-order. For a cut height of $g=1$, \preorder results in much smaller \pwmean[0] values than \inorder. The example in \autoref{fig:tree-layouts} suggests why: All the short edges in \inorder are near the bottom of the tree, where the contribution to \pwmean[0] is minimal. \comment{
As expected, \inorder results in much higher pointer-based search times than \preorder. 
}
However, this behavior changes as we increase the cut height, and at some point, in-order layouts are better than pre-order layouts. 
When the cut is approximately near halfway down the tree, in-order layouts such as \ivl result in much smaller \pwmean[0] values than pre-order layouts such as \pvl. As we increase the cut height all the way to $g=h-1$, we observe that the in-order version of the breadth-first layout \inbreadth continues to be better than the pre-order version \prebreadth.

\begin{table*}[t]
\centering
\begin{tabular}{c|c||c|c|c|c}
&
{\multirow{2}{*}{Cut height $g$}} &
{Pre-order layouts} &
\multicolumn{2}{c|}{Hybrid layouts} &
{In-order layouts} \\
\cline{3-6}
& & $\PR_\infty$ & $\IN_\infty$ & $\IN_2$ & $\IN_1$\\
\hline\hline
 Depth-first & $1$ & \preorder (${\PR_\infty^1}$) & \minwla ($\IN_\infty^1$) & \minep ($\IN_2^1$) & \inorder ($\IN_1^1$)\\
\hline
 Other & & \bender ($\PR^{h - 2^{\CEIL{\log_2(h/2)}}}_\infty$) & & \minwep ($\alt{\IN}_2^{opt}$)&\\
\hline
\comment{
 Optimal & & & & \approxwep ($\alt{\IN}_2^{g_{apx}}$), \minwep ($\alt{\IN}_2^{opt}$)&\\
\hline
}
 \multirow{2}{*}{\vEB} &
 \multirow{2}{*}{${\FLOOR{h/2}}$} 
 & \pvl ($\PR_\infty^{\FLOOR{h/2}}$) & & & \ivl ($\IN_1^{\FLOOR{h/2}}$)\\
\cline{3-6}
 & & \pvla ($\alt{\PR}_\infty^{\FLOOR{h/2}}$) & & \halfwep ($\alt{\IN}_2^{\FLOOR{h/2}}$) & \ivla ($\alt{\IN}_1^{\FLOOR{h/2}}$)\\
\hline
 Breadth-first & $h-1$ & \prebreadth ($\PR_*^{h-1}$) & & & \inbreadth ($\IN^{h-1}_*$)\\
\end{tabular}
\caption{
Nomenclature for \HLs.  The table summarizes the layouts discussed in the text, organized by cut height (rows) and subtree ordering (columns). The cut height function $g^\opt$ for \minwep is described in \autoref{sec:cut-hts}. The 
wild-card $*$ indicates that a particular parameter is not relevant.
}
\label{tbl:notation}
\end{table*}

\subsection{The computational cost of layouts: Pointer-less searches}

Based on explicit pointer-based search times, we have shown that \minwep is a cache-oblivious layout with almost $20\%$ improvement in performance when compared to the best in the literature, represented by \pvl. However, \minwep is a more complex layout than \pvl.
\comment{
and to ensure that the performance numbers are not affected by the time taken to compute the position of a \node in the layout, we have presented explicit pointer-based times so far. 
}
The natural question therefore is: If we considered implicit, pointer-less search times, would \minwep still compare favorably with \pvl? In \cite{Brodal:COSearch-vEBLayout}, the authors showed that for small tree heights, even layouts that have poor cache-performance such as \inorder and \prebreadth perform better than \pvl in implicit search, simply because it is trivial to compute the position of a \node in such layouts.

To understand the trade-offs involved, we first measured the time taken to compute the index of child nodes in a pointer-less search by excluding all memory accesses.%
\footnote{We achieved this by storing the keys $\{1, \ldots, |V|\}$ in the tree, allowing them to be easily inferred without lookup via their in-order index.}
\autoref{lst:code} lists the code that takes the \prebreadth index for a \node and computes its corresponding \minwep index.  This code needs to be executed for every transition in the search tree.  Here the depth (level) $d = \FLOOR{\log_2 i}$ of the node is maintained together with $i$ along the search path from the root.
Observe that one of the two functions in this code segment (\texttt{partition}) calculates the cut height for any pre-order subtree. 

In \autoref{fig:cut}, we see that the index computation time is almost constant for simple layouts (\inorder, \preorder, \inbreadth, and \prebreadth). The slow increase merely stems from the longer search paths as the height of the tree is increased. Furthermore, \minwep's index computation time is usually $4$ times that of the simple layouts. Comparing \minwep with the \vEBls (\ivl, \pvl, \bender, \halfwep) is more interesting. 
Not surprisingly, \halfwep performs worse than \ivl on this metric (around $20\%$ worse), since it is a more complex layout. Observe that \pvl performs better than \ivl, and by almost $50\%$. It turns out that the index can be computed more quickly within a pre-order subtree, since one does not need to keep track of left and right, and also because some other optimizations unique to pre-order layouts can be performed. This observation is key, since it allows us to compute the index for \minwep in $30\%$ less time than \halfwep, which is very similar at first glance. This is because we can set $g_I^\opt(h)=1$, as shown in \autoref{sec:cut-hts}, reducing the computational burden significantly by converting any in-order computation to a pre-order computation. As a result of this optimization, mean index computation times for \minwep are also about $20\%$ less than those of \ivl. Finally, observe that index computations take almost $60\%$ more time in \bender compared to \pvl, because of the additional time spent computing \bender's complex cut heights.

\autoref{fig:cut} also presents our results on implicit, pointer-less search times. One can think of these as a combination of the index computation times (which do not include memory accesses) and explicit search times (which include memory accesses, but avoid index computations using pointers). We see that for the more complex layouts, the implicit search times correlate very well with the index computation times. This is because of the relatively fast memory access times; if we added disk or even flash to the memory hierarchy, we would expect the relative order among the implicit times to be similar to the explicit times. 
The only perceptible difference in our experiments is that the pre-order layouts (\pvl and \bender) perform slightly worse, since they perform almost $20\%$ worse on the explicit search times. 
Among the simpler layouts, the implicit search times diverge significantly from the index computation times due to their poor memory access times. For tress of height $28$, \inorder already performs worse than \pvl, and we expect all of the simpler layouts to perform worse than \minwep as the height of the tree increases beyond $32$.

\subsection{Experimental Setup}
\label{sec:setup}

Our experiments were run on a single core of a
dual-socket 6-core 2.80~GHz Intel Xeon X5660 (Westmere-EP) processor with
96~GB of 3x DDR3-1333 RAM split over two 48~GB NUMA memory banks,
12~MB 16-way per-socket shared L3 cache,
256~KB 8-way L2 cache,
and 32~KB 8-way L1 data cache.
All three caches use 64-byte cache lines.
To reduce noise in the timing measurements, we computed the median time
of 15 runs.  Each run searches for (up to) 10 million randomly selected nodes.
We counted the number of L1 and L2 cache misses incurred in memory accesses
to the binary tree (stored as a linear array) using \texttt{valgrind-3.5.0}.
We also repeated our experiments on different architectures,
from powerful workstations to laptops, and observed similar results.

\todo{
Only search time exeriments here. since everything else has already been demonstrated?

No need to compare with DFS,etc, at this point. just i/p-VEB, and minWEP?

Cache Miss experiments.
Search time experiments. (Discuss trade-offs in calculation, compare with vEB.) 
}

\section{Conclusions}

In this paper, we present \minwep, a new layout for cache-oblivious search trees that outperforms layouts used in the literature by almost $20\%$. Using a general framework of \HLs, we showed that \minwep minimizes a new locality measure (\pwmean[0], which represents the \WEP) that correlates very well with cache-misses in a multi-level cache hierarchy. 
All widely used cache-oblivious versions of search trees rely on \vEBls, which are shown to be a special case of \HLs. Therefore, we suggest that the performance of all these data structures can be easily improved by switching to a layout that is derived from \minwep.

While enumerating all possible orderings for small trees, we noticed that the optimal \pwmean[0] value is sometimes obtained by layouts that do not place the top subtree at one end or in the middle of the bottom subtrees. This implies that \RLs do not necessarily optimize \pwmean[0]. One direction of future study is to generalize the notion of \RLs to include such \HLs, and to construct unrestricted layouts that optimize \pwmean[0]. We would also like to prove that, at least among all \RLs, \minep and \minwep minimize \pmean[0] and \pwmean[0], respectively, since we believe this is true based on our extensive empirical study.
\note{These seem like contradictory statements.  If \minwep is a \HL and we think it minimizes \pwmean[0], then we could not have noticed a better one. Agreed, changed the text.}

\todo{
Finding optimal layout for WEP.
Extending memory transfer bounds.
Extensions to streaming etc can just use our vEB layout.
}

\bibliographystyle{IEEEtran}
\bibliography{paper,minwep}


\begin{lstlisting}[float,caption={Breadth-first to \minwep index translation.},label=lst:code]
uint partition(uint h) {                 // height h of tree to partition
    return h <= 5 ? 1 : (h - 1) / 2;     // return top subtree height
}

uint index(uint i, uint d, uint h) {     // BF index i, node depth d, tree height h
    uint p = 1 << --h;                   // MinWEP index being computed
    while (d) {                          // iterate until node is root of subtree
        uint q = (i >> --d) & 1;         // initial offset (pre: q=1; post: q=0)
        uint r = q - 1;                  // bit reversal (pre: r=0; post: r=~0)
        i ^= r;                          // post-order is reversal of pre-order
        while (d) {                      // iterate until node is root of subtree
            uint g = partition(h);       // top subtree height
            if (d < g) {                 // is node in top subtree?
                h = g;                   // set height to top subtree height
                i = ~i;                  // alternate left/right ordering
            } else {                     // node is in bottom subtree
                h -= g;                  // bottom subtree height
                d -= g;                  // depth within bottom subtree
                uint m = (1 << g) - 1;   // number of nodes in top subtree
                q += m;                  // advance past top subtree
                uint k = (i >> d) & m;   // subtree number (pre: k=0; in: 1<=k<=m)
                if (k) {                 // in in-order subtree?
                    q += (k << h) - k;   // advance past k bottom subtrees
                    q += (1 << --h) - 1; // advance to root of in-order subtree
                    break;               // transition to in-order case
        }   }   }
        i ^= r;                          // restore i if post-order
        q ^= r;                          // negate offset if post-order
        p += q;                          // advance to smaller in-order subtree
    }
    return p;                            // return MinWEP index
}
\end{lstlisting}

\input{treeplots}

\clearpage

\input{appendix}

\end{document}

%% file: macros.tex
\newcommand{\notetext}[1]{%
  \raggedright\tiny\sffamily\bfseries{#1}%
}
\newcommand{\sidenote}[1]{%
  \ifinner%
    \smash{%
      \raggedright%
      \hspace*{\textwidth}%
      \hspace*{\marginparsep}%
      \parbox[t]{\marginparwidth}{\notetext{#1}}%
    }%
    \\[-0.5\bs]%
  \else%
    \marginpar{\notetext{#1}}%
  \fi%
}

\newcommand{\note}[1]{}
\newcommand{\C}[1]{}
\newcommand{\comment}[1]{}

\newcommand{\node}{node\xspace}
\newcommand{\nodes}{nodes\xspace}

\newcommand{\vEB}{van Emde Boas\xspace}
\newcommand{\vEBl}{van Emde Boas layout\xspace}
\newcommand{\vEBls}{van Emde Boas layouts\xspace}
\newcommand{\HL}{Hierarchical Layout\xspace}
\newcommand{\HLs}{Hierarchical Layouts\xspace}
\newcommand{\RL}{Recursive Layout\xspace}
\newcommand{\RLs}{Recursive Layouts\xspace}
\newcommand{\pvl}{\textsc{Pre-vEB}\xspace}
\newcommand{\pvla}{\textsc{Pre-vEBa}\xspace}
\newcommand{\ivl}{\textsc{In-vEB}\xspace}
\newcommand{\ivla}{\textsc{In-vEBa}\xspace}
\newcommand{\bender}{\textsc{Bender}\xspace}
\newcommand{\preorder}{\textsc{Pre-Order}\xspace}
\newcommand{\inorder}{\textsc{In-Order}\xspace}

\newcommand{\prebreadth}{\textsc{Pre-Breadth}\xspace}
\newcommand{\inbreadth}{\textsc{In-Breadth}\xspace}
\newcommand{\WEP}{Weighted Edge Product\xspace}

\newcommand{\PR}{\ensuremath{\mathcal{P}}\xspace}
\newcommand{\IN}{\ensuremath{\mathcal{I}}\xspace}
\newcommand{\alt}[1]{\widetilde{#1}}
\newcommand{\opt}{\ensuremath{\mathit{opt}}\xspace}

\newcommand{\CEIL}[1]{\ensuremath{\lceil #1 \rceil}\xspace}
\newcommand{\FLOOR}[1]{\ensuremath{\lfloor #1 \rfloor}\xspace}

\newcommand{\eg}{{\it e.g.}\xspace}
\newcommand{\ie}{{\it i.e.}\xspace}

\newcommand{\pmean}[1][p]{\ensuremath{\mu_{#1}}\xspace}
\newcommand{\pwmean}[1][p]{\ensuremath{\nu_{#1}}\xspace}
\newcommand{\PB}{\ensuremath{\beta}\xspace}
\newcommand{\pos}[1][]{%
  \def\tmp{#1}%
  \ifx\tmp\empty%
    \ensuremath{\varphi}\xspace%
  \else%
    \ensuremath{\varphi(#1)}\xspace%
  \fi%
}
\newcommand{\weight}[1][]{\ensuremath{w_{#1}}\xspace}

\newcommand{\prob}[1][]{\ensuremath{p_{#1}}\xspace}
\newcommand{\minbw}{\textsc{MinBW}\xspace}
\newcommand{\minla}{\textsc{MinLA}\xspace}
\newcommand{\minwla}{\textsc{MinWLA}\xspace}
\newcommand{\minep}{\textsc{MinEP}\xspace}
\newcommand{\minwep}{\textsc{MinWEP}\xspace}
\newcommand{\halfwep}{\textsc{HalfWEP}\xspace}
\newcommand{\approxwep}{\textsc{ApproxWEP}\xspace}

\newcommand{\betaomega}[1][p]{\mbox{\ensuremath{\beta\text{-}\Omega}}\xspace}

\newcommand{\nodeset}[1][]{\ensuremath{V_{#1}}\xspace}
\newcommand{\edges}[1][]{\ensuremath{E_{#1}}\xspace}

\newcommand{\len}[1][]{\ensuremath{\ell_{#1}}\xspace}
\newcommand{\treelen}[2]{\ensuremath{\ell^{#1}_{#2}}\xspace}
\newcommand{\csize}[1][]{\ensuremath{N}\xspace}

\newcommand{\acmr}[1][]{%
  \def\acmrarg{{#1}}%
  \acmri%
}
\newcommand{\acmri}[1][]{%
  \def\acmrargii{#1}%
  \ifx\acmrargii\empty%
    \ensuremath{M_{\acmrarg}}\xspace%
  \else%
    \ensuremath{M_{\acmrarg}(#1)}\xspace%
  \fi%
}

\newlength{\boxwidth}

%% file: abstract.tex
This paper proposes a general framework for generating cache-oblivious layouts for binary search trees. A cache-oblivious layout attempts to minimize cache misses on any hierarchical memory, independent of the number of memory levels and attributes at each level such as cache size, line size, and replacement policy. Recursively partitioning a tree into contiguous subtrees and prescribing an ordering amongst the subtrees, \emph{Hierarchical Layouts} generalize many commonly used layouts for trees such as in-order, pre-order and breadth-first. They also generalize the various flavors of the van Emde Boas layout, which have previously been used as cache-oblivious layouts. Hierarchical Layouts thus unify all previous attempts at deriving layouts for search trees.

The paper then derives a new locality measure (the Weighted Edge Product) that mimics the probability of cache misses at multiple levels, and shows that layouts that reduce this measure perform better. We analyze the various degrees of freedom in the construction of Hierarchical Layouts, and investigate the relative effect of each of these decisions in the construction of cache-oblivious layouts. Optimizing the Weighted Edge Product for complete binary search trees, we introduce the \minwep layout, and show that it outperforms previously used cache-oblivious layouts by almost 20\%.

%% file: measure.tex
We have seen how the percentage of block transitions provides a
quality measure for a layout given a particular cache block size \csize.
We now remove this dependence on block size and derive a simple
measure of locality for graph orderings in a
cache-oblivious sense, \ie with no knowledge of cache and line size.
Continuing the discussion in
\autoref{sec:blocktrans},
we here generalize the measure presented in~\cite{Yoon06} to
weighted graphs.


\C{
As
in~\cite{Yoon05,Yoon06,Tchiboukdjian10}, we model data elements as
vertices $V$ in a graph $G(V, E)$, with an undirected edge
indicating a nonzero likelihood that its two vertices be accessed in
succession, or more generally a desire to store those two vertices close
together in memory.  We call such a graph an \emph{affinity graph}.

The affinity between $i$ and $j$ may be expressed in terms of a
weight $\weight[ij] = \weight[ji] > 0$.  Let $A$ denote the
matrix of affinities, such that $a_{ij} = w_{ij}$ if $ij \in E$
and $a_{ij} = 0$ otherwise. 
 We model data accesses as a Markov
chain random walk on $G$ with transition matrix $P = D^{-1} A$,
where $D$ is the diagonal matrix with $d_{ii} = \sum_j a_{ij}$.
If $G$ is strongly connected,%
\footnote{Because each component of a
graph may be laid out independently, we assume that $G$ is strongly
connected.}
then it is well-known that the probability $\Pr(X_t = i, X_{t+1} = j)$
of being in state $i$ and transitioning to state $j$ equals
$\frac{\weight[ij]}{W}$, where $W = \sum_{ij \in E} \weight[ij]$.
In other words, the probability of accessing two data elements in
succession is proportional to the weight of the edge connecting them.

Consider a cache consisting of a single block that can hold \csize
data elements and is backed by a larger memory consisting of several
such blocks.  (In practice caches tend to hold more than one block,
but that would unnecessarily complicate our derivation.)
Let $i$ and $j$ be data elements stored at $\pos[i]$ and
$\pos[j]$ in blocks $B(i)$ and $B(j)$, respectively, and separated by
$\len[ij] = |\pos[i] - \pos[j]|$ on linear storage.  Without loss of
generality, assume $\pos[i] < \pos[j]$.  Suppose $i$ is accessed
first, bringing $B(i)$ into the cache.  We wish to estimate the
probability of a cache miss when $j$ is accessed.  Clearly, if
$\ell \geq \csize$, then a cache miss is inevitable, since then
$B(i) \neq B(j)$.  When $\ell < \csize$, the likelihood of a cache miss
depends on the position of $i$ within $B(i)$, \ie where the upper
boundary of $B(i)$ lies with respect to $\{\pos[i], \ldots, \pos[j]\}$.
In absence of further information, we will assume that the position
of $i$ within $B(i)$ is distributed uniformly.\footnote{Even in
practice, modern operating systems allocate memory blocks with
nearly arbitrary alignment.}
Hence, there are $\ell$ out of $\csize$ possible alignments
that partition $i$ and $j$ into separate blocks, and the
probability of a cache miss occurring when $j$ is accessed is
\begin{equation}
  \acmr[\csize][\len] =
  \begin{cases}
        \frac{\len}{\csize} & \text{if $\len < \csize$} \\
        1 & \text{otherwise}
  \end{cases}
\end{equation}
}

The observation underlying our cache-oblivious measure is that most block-based caches 
employed in current computer architectures are hierarchical and
nested, with a roughly geometric progression in
size.
\C{
\footnote{For instance, the memory hierarchy for the computer
used in this paper consists of
$2^{16} = 64~\text{KB}$ L1 cache,
$2^{18} = 256~\text{KB}$ L2 cache,
$2^{23} = 8~\text{MB}$ L3 cache, 
$2^{34} = 16~\text{GB}$ of RAM, and
$2^{40} = 1~\text{TB}$ of disk.}
}
That is, we may write $\csize = b^k$ for some base $b$ (usually $b = 2$)
and positive integer $k$. 
We then estimate the total
number of cache misses for all $k$ for a particular edge length \len as
\begin{equation}
\begin{split}
  \acmr[][\len] & = \sum_{k=1}^\infty \acmr[b^k][\len]
                  = \sum_{k=1}^{\lfloor \log_b \len \rfloor} 1 +
                    \sum_{k=\lfloor \log_b \len \rfloor + 1}^\infty \frac{\len}{b^k} \\
                & = \lfloor \log_b \len \rfloor +
                    \len \frac{b^{-\lfloor \log_b \len \rfloor}}{b - 1}
\end{split}
\end{equation}
We note that when \len is an exact power of $b$, $\acmr[][\len]$ simplifies
to $\log_b \len + \frac{1}{b - 1}$; otherwise $\acmr[][\len]$ increases
monotonically with $\len$.  Our primary goal is not to estimate the
exact number of cache misses incurred, but rather to assign a relative ``cost''
as a function of edge length \len.  We may thus ignore the value of $b$ (since
it affects only the slope of $\acmr$) and the constant term independent of
$\len$, and arrive at the approximation
\begin{equation}
  \acmr[][\len] \approx \log \len
\end{equation}
Intuitively, $\log \len[ij]$ measures the number of blocks smaller than \len[ij] that cannot hold both $i$ and $j$, and thus captures the expected number of block transitions and cache misses
associated with \len[ij] in a memory hierarchy.
\comment{
The quality of this approximation is quite good,
as evidenced by
\autoref{fig:acmr-vs-log}.

\begin{figure}[tp]
\centering%
\subfloat[]{%
  \includegraphics[width=0.48\linewidth]{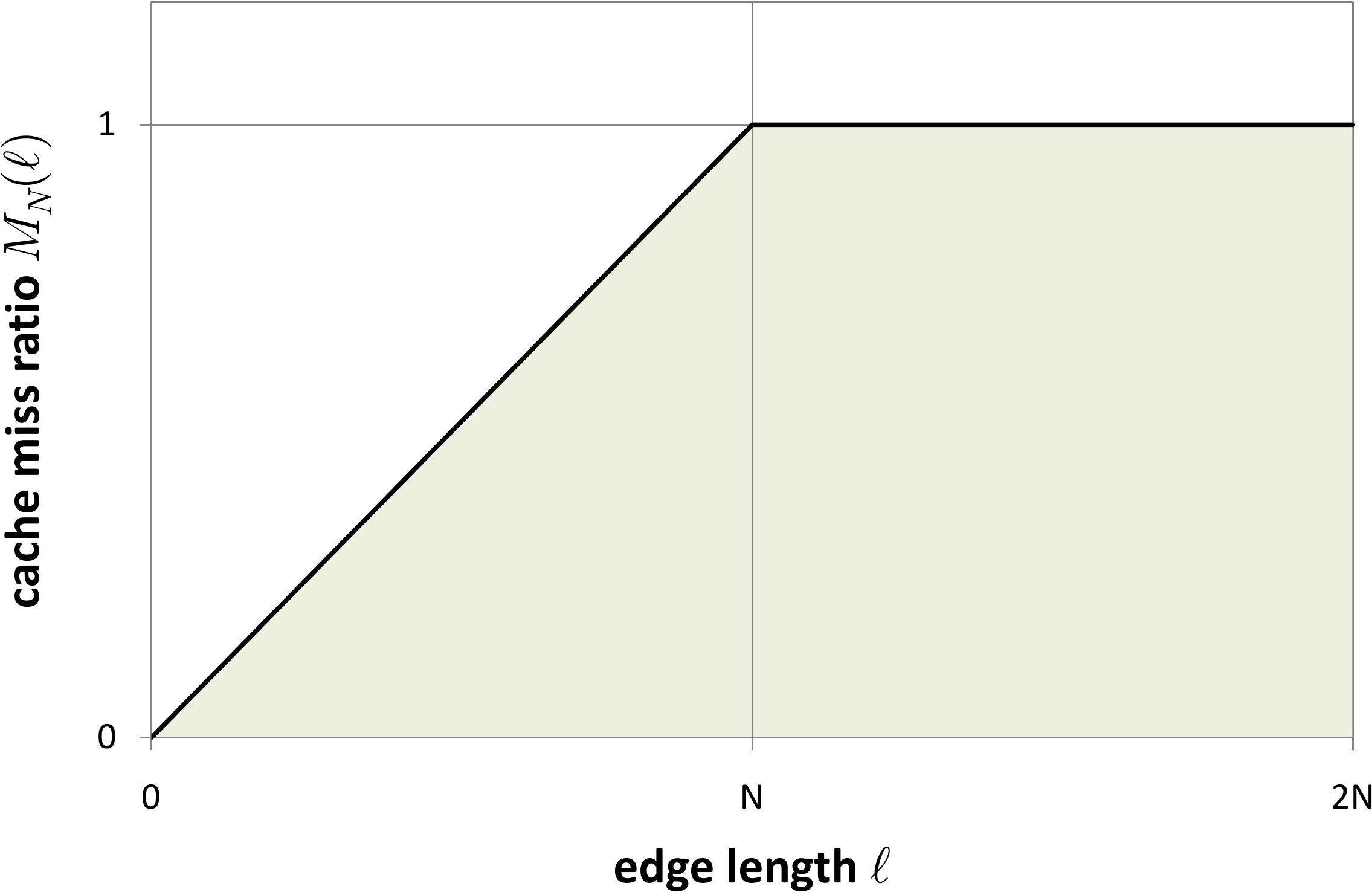}%
  \label{fig:acmr-vs-len}%
}%
\hfill%
\subfloat[]{%
  \includegraphics[width=0.48\linewidth]{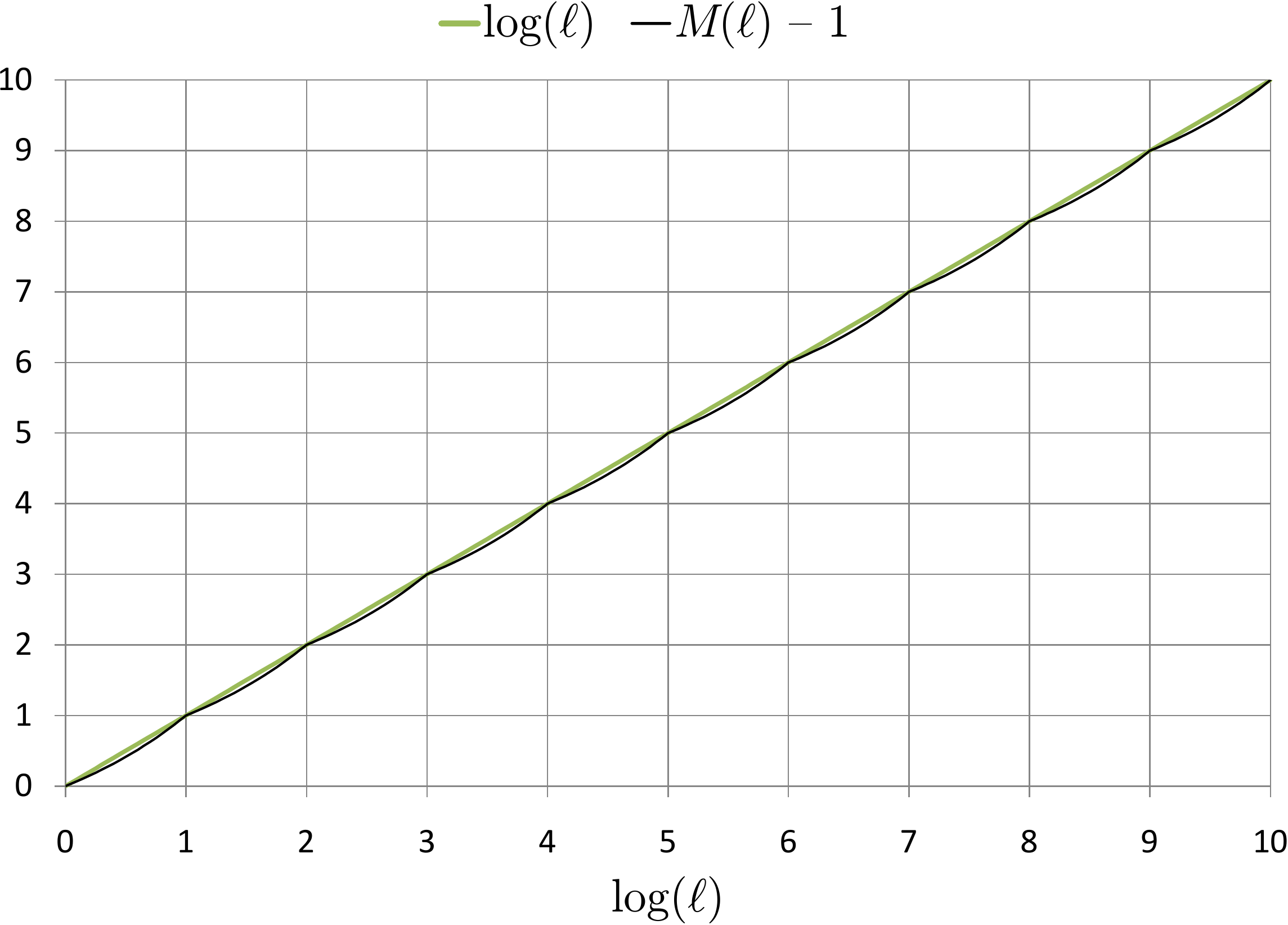}%
  \label{fig:acmr-vs-log}%
}%
\caption{
  (a)~The cache miss ratio \acmr[\csize] as a function of edge length is linear
  up to the cache block size \csize.
  (b)~The \emph{average cache miss ratio} ($\acmr$) is a measure of the expected
  number of cache misses across a memory hierarchy.  This plot shows that
  the \acmr associated with accessing two data elements is strongly
  correlated with the logarithm of the distance \len between the elements.
}
\label{fig:acmr-vs-lenlog}
\end{figure}
}

Finally, if we consider all edges $E$ of an affinity graph, then
\begin{equation}
\begin{split}
  \acmr & = \sum_{k=1}^\infty \PB(b^k)
          = \sum_{k=1}^\infty \frac{1}{W} \sum_{ij \in E} \weight[ij] \acmr[b^k][{\len[ij]}] \\
        & = \frac{1}{W} \sum_{ij \in E} \weight[ij] \sum_{k=1}^\infty \acmr[b^k][{\len[ij]}]
          = \frac{1}{W} \sum_{ij \in E} \weight[ij] \acmr[][{\len[ij]}] \\
        & \approx \frac{1}{W} \sum_{ij \in E} \weight[ij] \log \len[ij]
          = \log \pwmean[0]
\end{split}
\label{eqn:acmr}
\end{equation}
gives the \emph{average cache miss ratio},
where $\pwmean[0]$ denotes the \emph{weighted edge product} functional
\begin{equation}
\begin{split}
  \pwmean[0]
  & = \exp\Biggl(
        \frac{1}{W} \sum_{ij \in E} \weight[ij] \log \len[ij]
      \Biggr)
    = \Biggl(
        \prod_{ij \in E} \len[ij]^{\weight[ij]}
      \Biggr)^{1/W}
\end{split}
\end{equation}
for a weighted graph.  
In other words, $\pwmean[0] \approx \exp(\acmr)$.  As a result, low values of
\pwmean[0] imply good cache utilization 
across the whole memory hierarchy.  As we shall see, this
expected behavior is observed also in practice, with layouts optimized
for \pwmean[0] having excellent locality properties.
(In the unweighted case, $\weight[ij] = 1$ and $W = |E|$. We denote the unweighted version of \pwmean[0] by \pmean[0].)
\comment{
As discussed in \autoref{sec:blocktrans}, the weights \weight[ij] are
given 
by the likelihood of visiting in succession \nodes $i$ and $j$ at
levels $d-1$ and $d$, respectively,
\ie 
by the likelihood of
searching for $j$ or one of its descendants.
}
We call the \RL that minimizes \pwmean[0] for
geometrically decreasing weights (as described in \autoref{sec:blocktrans})
the \minwep
(short for minimum weighted edge product) layout of the tree.
\comment{
This name derives from the fact that when $\weight[ij] = 1$, the
optimal layout minimizes the (unweighted) product of edge lengths.
}

\note{Point out why $\log \len$ is the right thing---a reduction at short
lengths matters more than the same at long lengths, as long edges are already
out of cache. Done.}

%% file: weighted.tex
\subsection{Other edge-based locality measures}

It is important to mention two other locality measures that
have been considered in the literature: the average edge
length, \pmean[1], and the maximum edge length, \pmean[\infty].
The small example in \autoref{fig:tree-layouts} 
includes the layouts \minla~\cite{Chung:MinLA} in
\autoref{fig:tree-minla}, which minimizes \pmean[1], and
\minbw~\cite{Heckmann:MinBW} in \autoref{fig:tree-minbw}, which
minimizes \pmean[\infty].  
Similar to \pwmean[0], which measures the weighted edge
length product, we may define the average weighted edge length \pwmean[1].
This figure also presents 
these four statistics
(\pwmean[0], \pwmean[1], \pmean[1], \pmean[\infty]) for 
all layouts discussed in this paper. 
From the discussion in \autoref{sec:blocktrans}, we observe that the
probability of a block transition for very large block sizes is given
by \pwmean[1], \ie, a weighted version of \pmean[1]. 

Based on an empirical study, we conjecture that among all \RLs \pwmean[1] is minimized by \minwla, the layout that
cuts at height $g=1$, arranges the outermost top subtree in-order, and
arranges every subsequent subtree pre-order.
In our nomenclature, \minwla is $\IN^1_\infty$ (see \autoref{tbl:notation}).
Restricting ourselves to \RLs with cut height $g=1$, \minwla provably minimizes \pwmean[1].
We delegate all proofs to the 
Appendix.
\begin{theorem}
\label{thm:minwla}
The \minwla layout minimizes \pwmean[1] among all \RLs with cut height $g=1$.
\end{theorem}

\begin{figure}[tp]
\centering%
\includegraphics[width=\columnwidth]{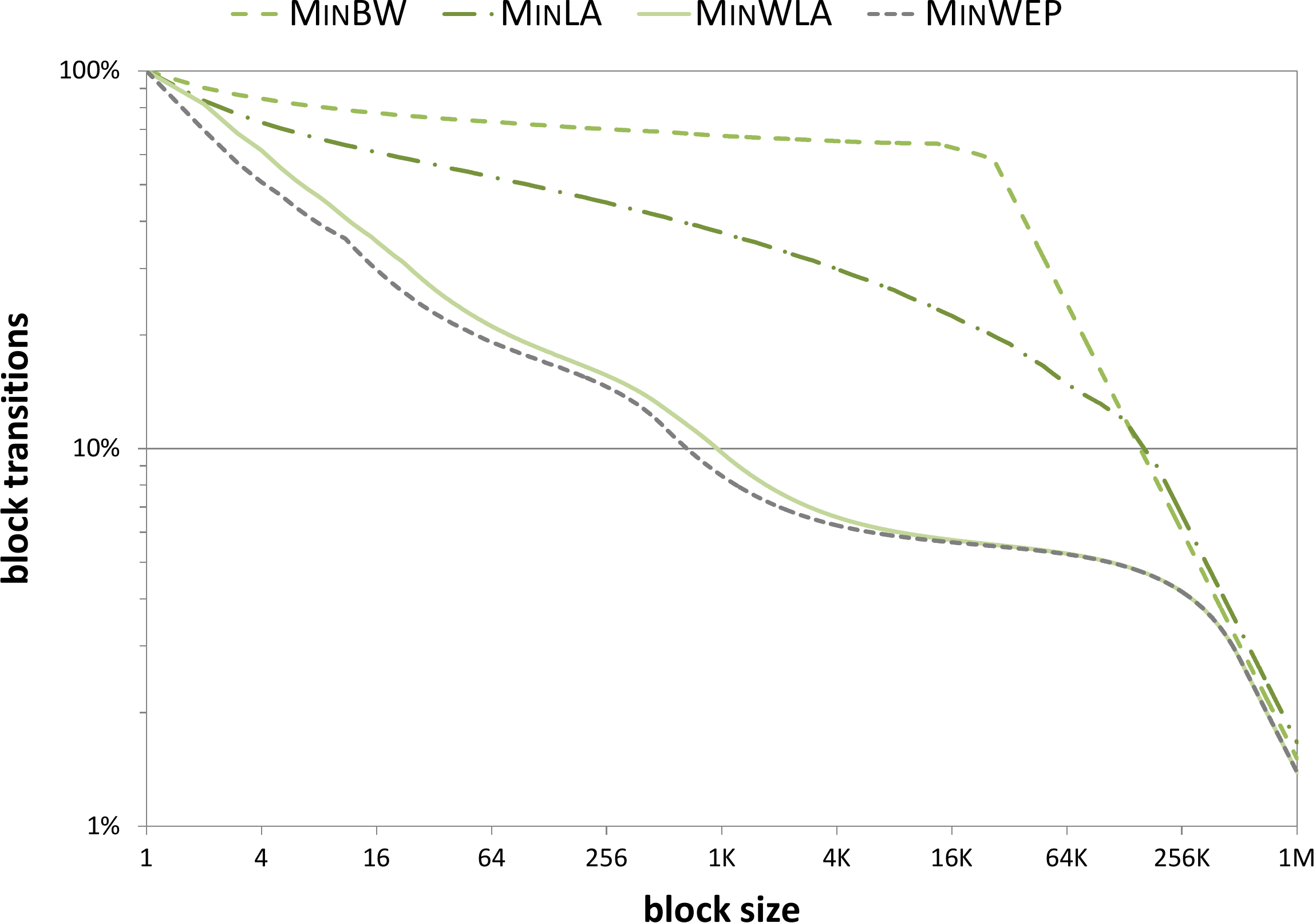}%
\caption{
  Block transitions for $h = 20$ for the layouts
  that minimize
  \pmean[\infty] (\textsc{BW}),
  \pmean[1] (\textsc{LA}),
  \pwmean[1] (\textsc{WLA}),
  \pwmean[0] (\textsc{WEP}).
}%
\label{fig:min-transitions}%
\end{figure}

Our experiments on block transitions
(\autoref{fig:min-transitions}), observed cache misses, and timings
indicate that these other layouts have significantly
worse locality than \minwep, lending support to our claim that
the weighted edge product (represented by \pwmean[0]) is the correct
measure to consider. 

In this paper, we present a succession of \HLs that reduce \pwmean[0],
and we see that these also tend to reduce \pwmean[1], \pmean[1], and
\pmean[\infty], suggesting that these might be good layouts in other
settings that benefit from better locality. 
\note{In fact, \cite{Safro11} minimizes \pwmean[0]. Added sentences.}
In \cite{Safro11}, the authors show that minimizing \pwmean[0]
results in compression-friendly layouts.
We note that minimizing \pwmean[0] is likely to result in high locality layouts 
for all graphs, and not just trees. 
For algorithms designed to minimize
\note{Have we defined \pmean[0]? Defined earlier.}
\pmean[0], \pwmean[0], \pmean[1], and \pmean[\infty], respectively, on general graphs,
see~\cite{Yoon06}, \cite{Safro11}, \cite{Safro09}, and \cite{Cuthill:minbw}.

%% file: treeplots.tex

\newcommand{\stats}[4]{\ensuremath{\pwmean[0] = #1, \pwmean[1] = #2, \pmean[1] = #3, \pmean[\infty] = #4}}

\begin{figure*}[tp]
\vspace*{-2ex}%
\subfloat[{\minwep = \minep: \stats{1.818}{4.063}{2.581}{23}}]{%
  \includegraphics[width=0.5\linewidth]{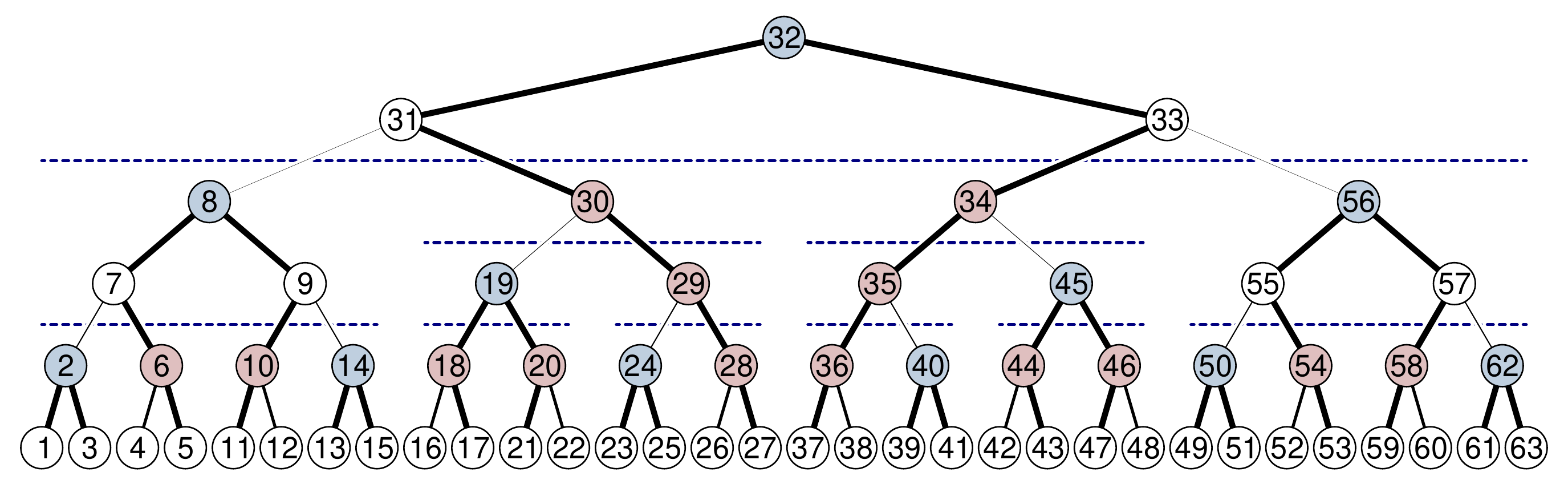}%
  \label{fig:tree-minwep}%
  \label{fig:tree-minep}%
}%
\subfloat[{\halfwep: \stats{1.823}{3.938}{3.097}{26}}]{%
  \includegraphics[width=0.5\linewidth]{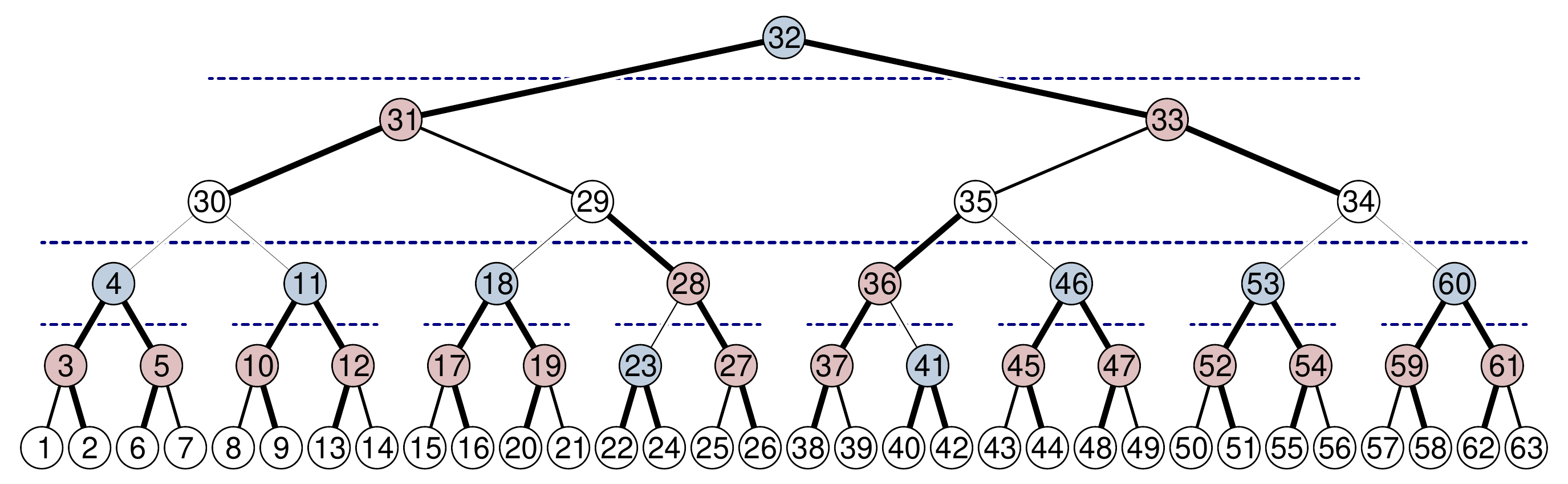}%
  \label{fig:tree-halfwep}%
}\\[-2ex]%
\subfloat[{\ivla: \stats{2.184}{4.300}{3.161}{27}}]{%
  \includegraphics[width=0.5\linewidth]{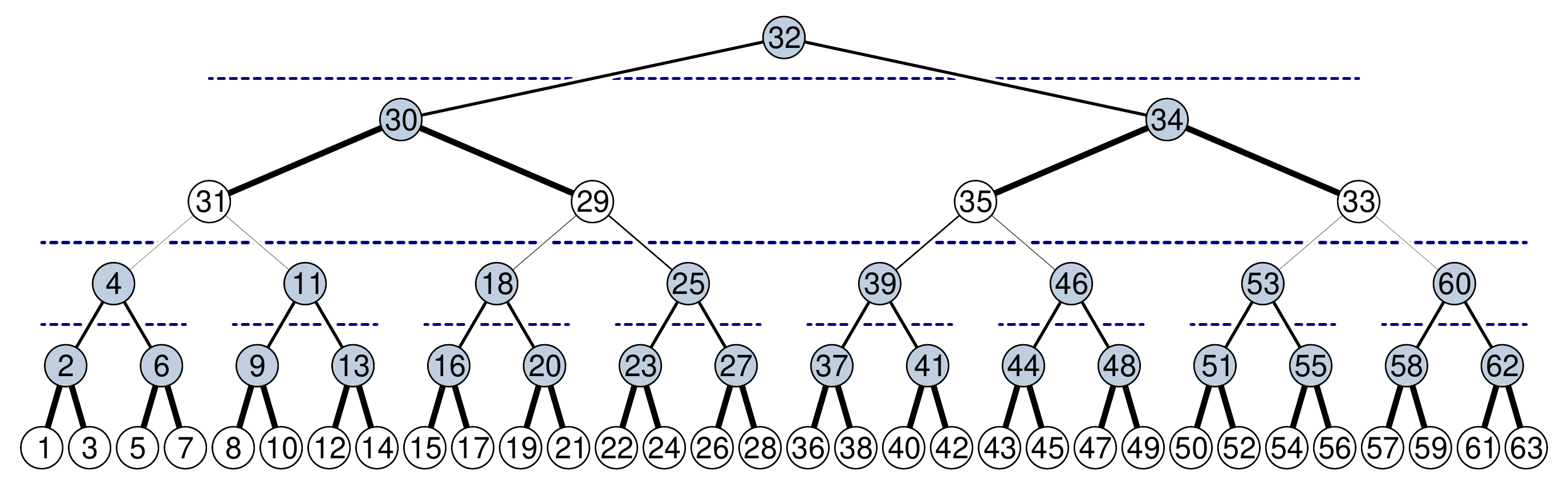}%
  \label{fig:tree-vebia}%
}%
\subfloat[{\pvla: \stats{2.691}{7.100}{5.145}{54}}]{%
  \includegraphics[width=0.5\linewidth]{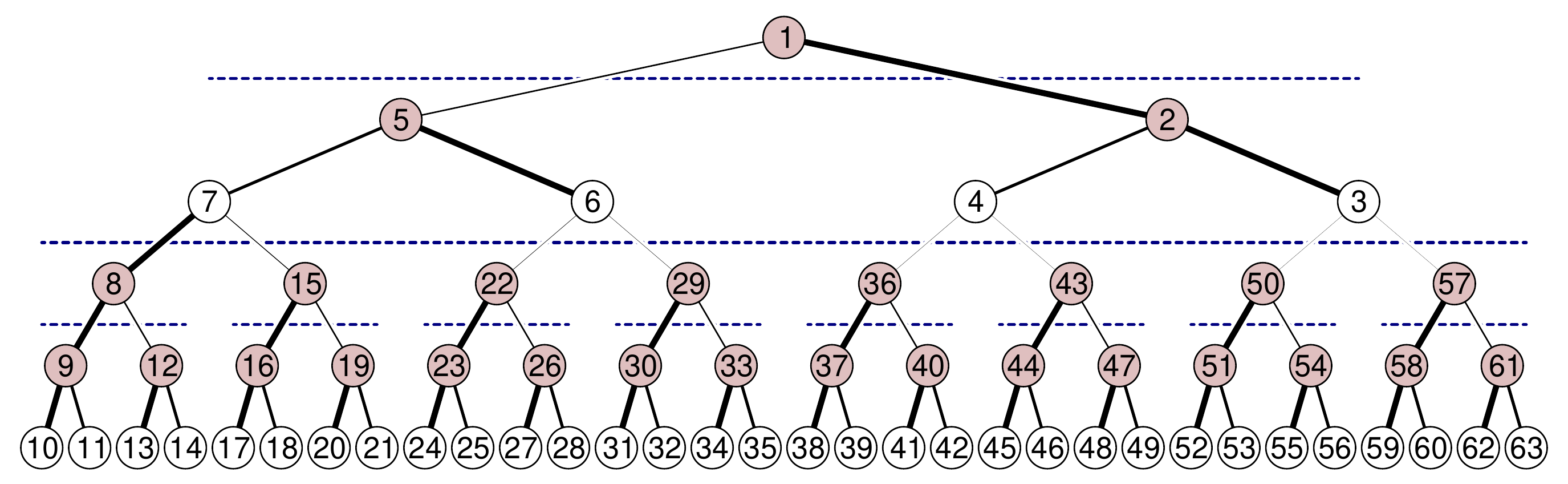}%
  \label{fig:tree-veboa}%
}\\[-2ex]%
\subfloat[{\ivl: \stats{2.227}{4.300}{3.161}{25}}]{%
  \includegraphics[width=0.5\linewidth]{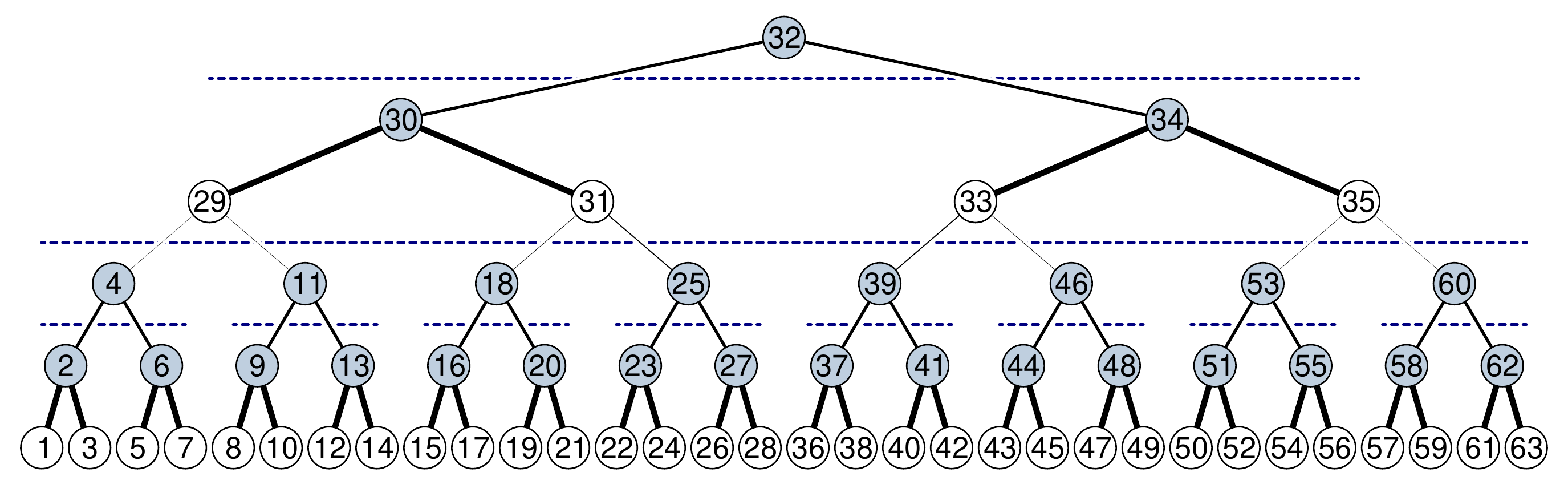}%
  \label{fig:tree-vebi}%
}%
\subfloat[{\pvl: \stats{2.824}{7.100}{5.145}{50}}]{%
  \includegraphics[width=0.5\linewidth]{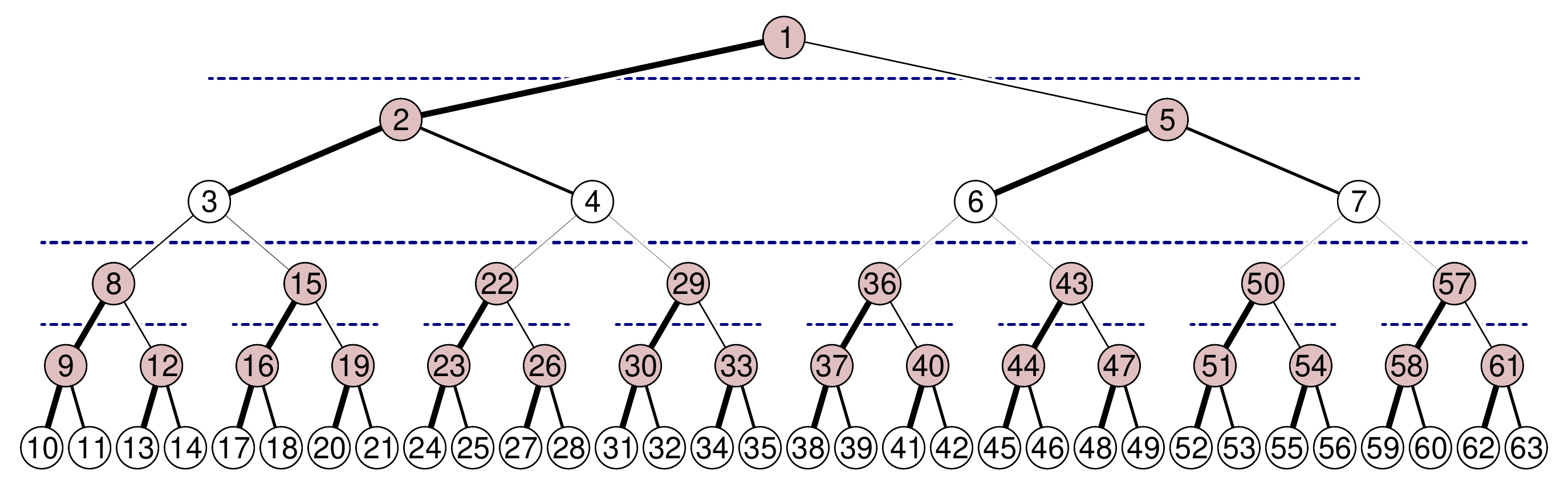}%
  \label{fig:tree-vebo}%
}\\[-2ex]%
\subfloat[{\inorder: \stats{4.000}{6.200}{2.581}{16}}]{%
  \includegraphics[width=0.5\linewidth]{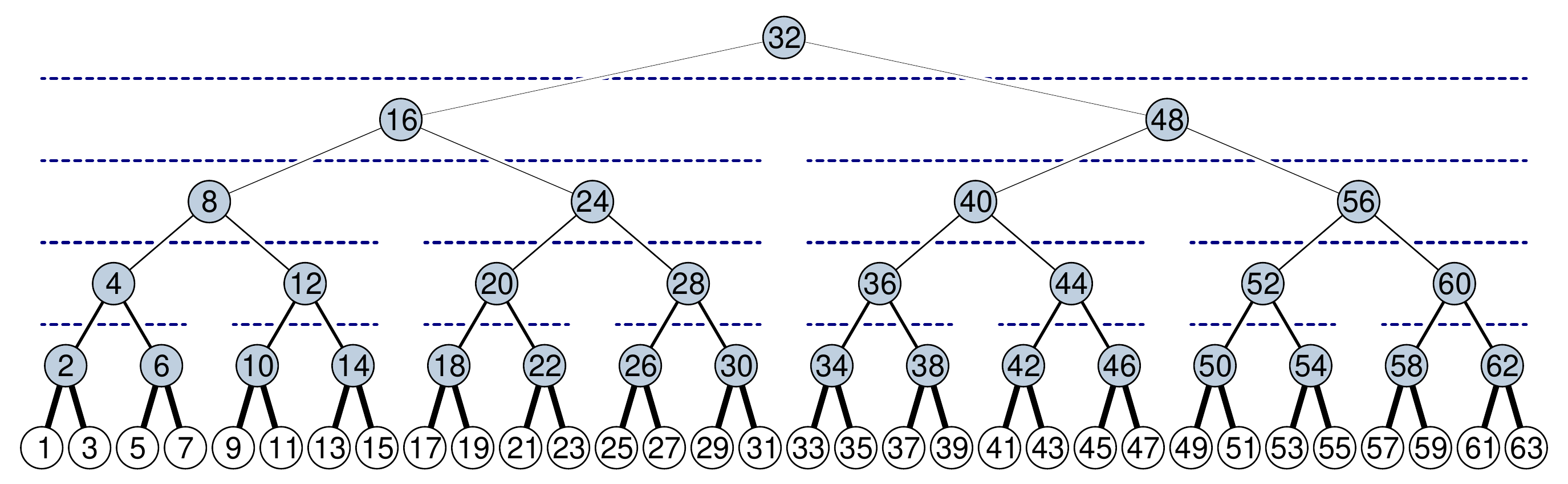}\vspace*{-1ex}%
  \label{fig:tree-inorder}%
}%
\subfloat[{\preorder: \stats{2.828}{6.700}{3.081}{32}}]{%
  \includegraphics[width=0.5\linewidth]{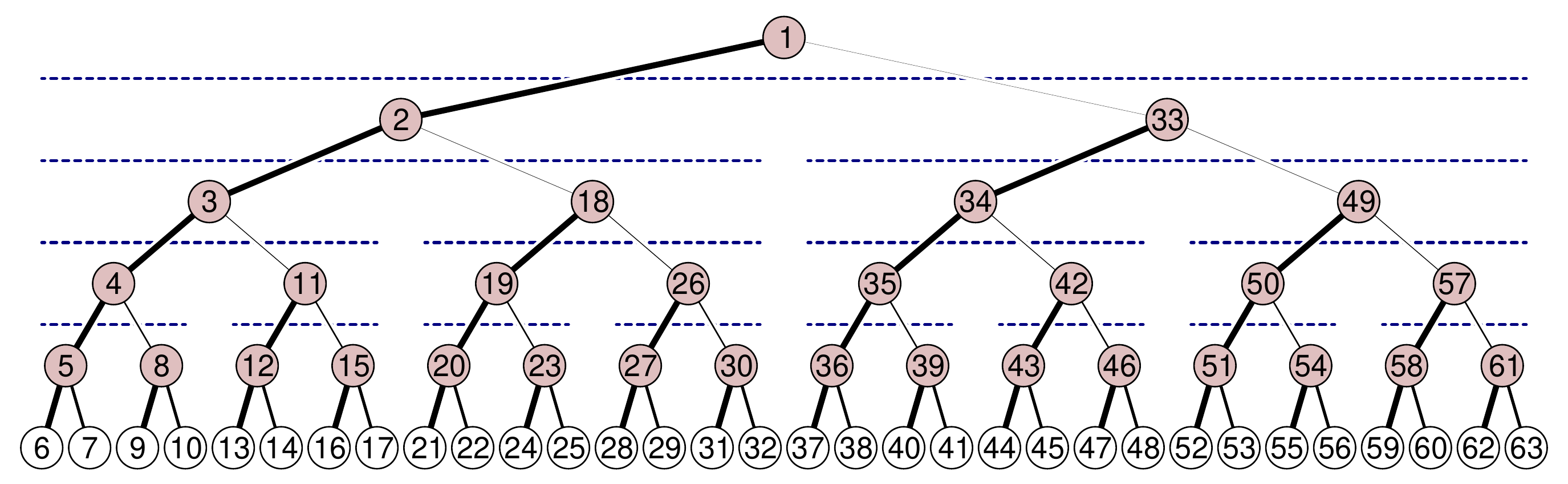}%
  \label{fig:tree-preorder}%
}\\[-2ex]%
\subfloat[{\inbreadth: \stats{3.096}{4.700}{8.258}{16}}]{%
  \includegraphics[width=0.5\linewidth]{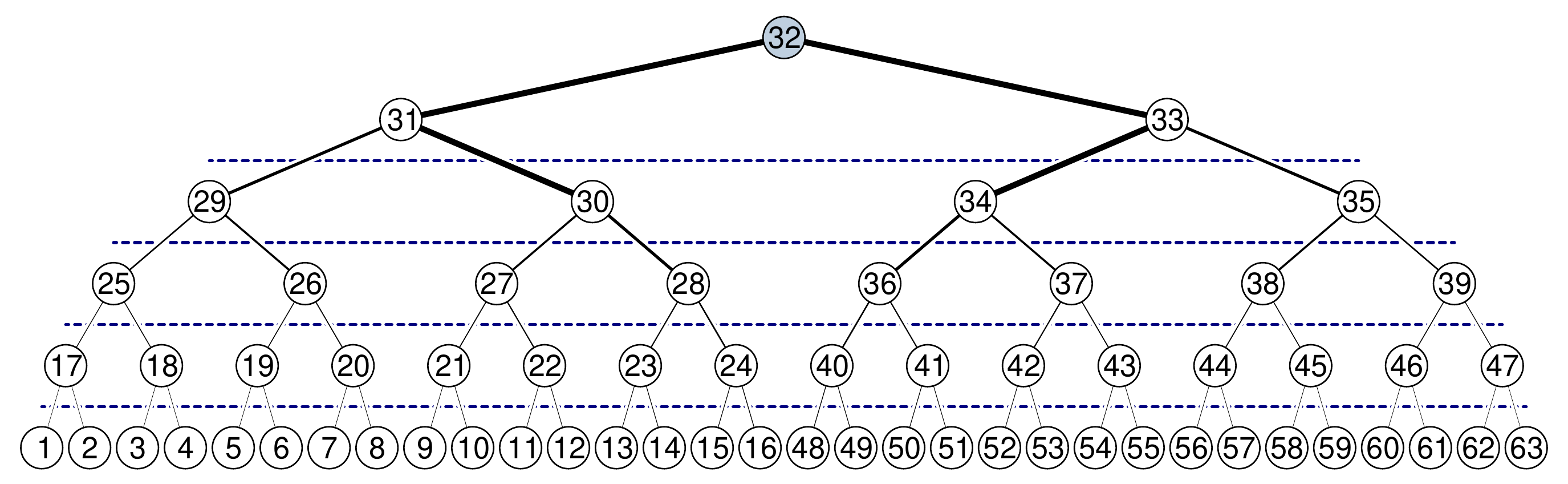}%
  \label{fig:tree-inbreadth}%
}%
\subfloat[{\prebreadth: \stats{5.824}{9.300}{16.500}{32}}]{%
  \includegraphics[width=0.5\linewidth]{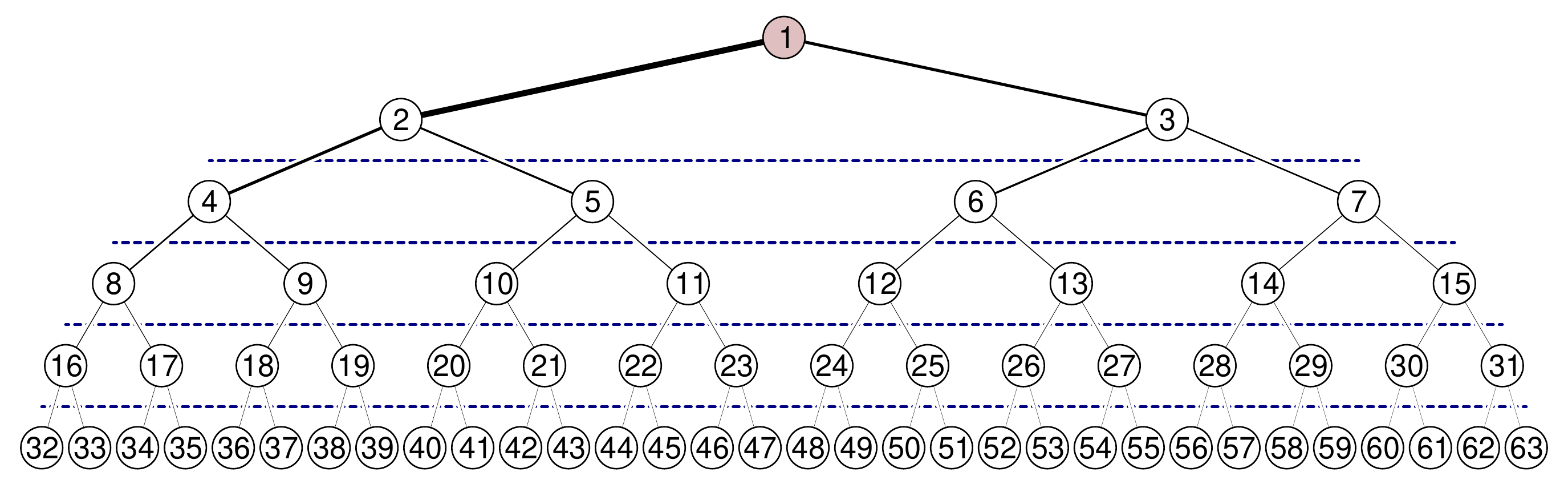}%
  \label{fig:tree-breadth}%
}\\[-2ex]%
\subfloat[{\minwla: \stats{2.000}{3.600}{2.581}{16}}]{%
  \includegraphics[width=0.5\linewidth]{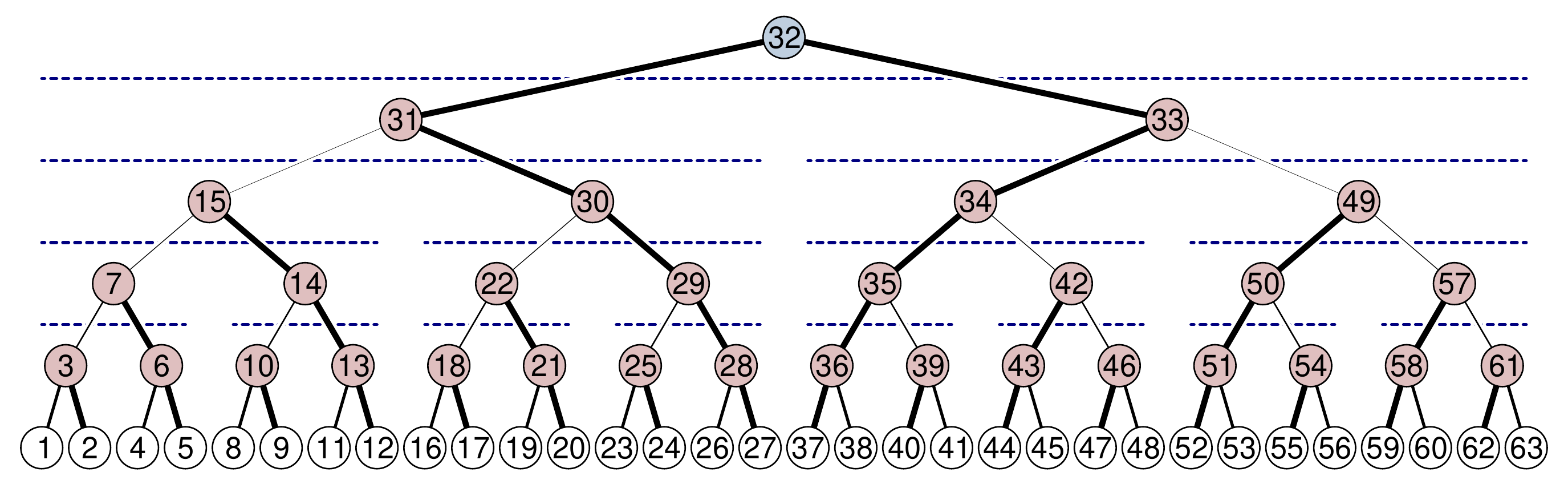}%
  \label{fig:tree-minwla}%
}%
\subfloat[{\bender: \stats{2.930}{6.900}{4.113}{46}}]{%
  \includegraphics[width=0.5\linewidth]{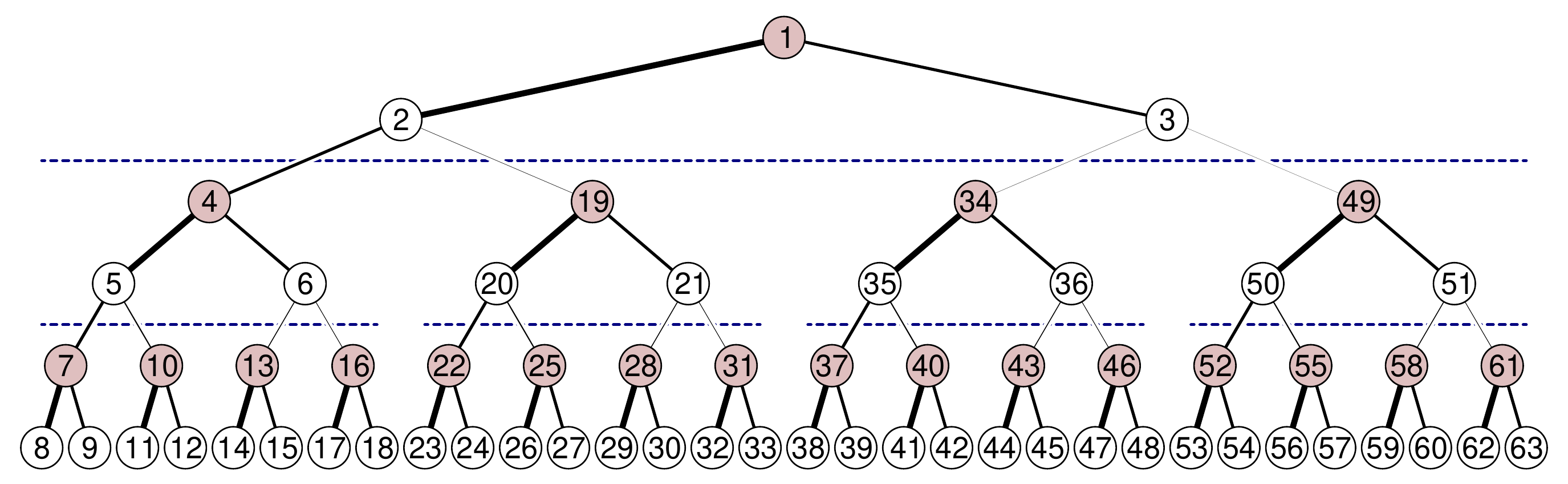}%
  \label{fig:tree-bender}%
}\\[-2ex]%
\subfloat[{\minla: \stats{2.753}{4.175}{2.323}{12}}]{%
  \includegraphics[width=0.5\linewidth]{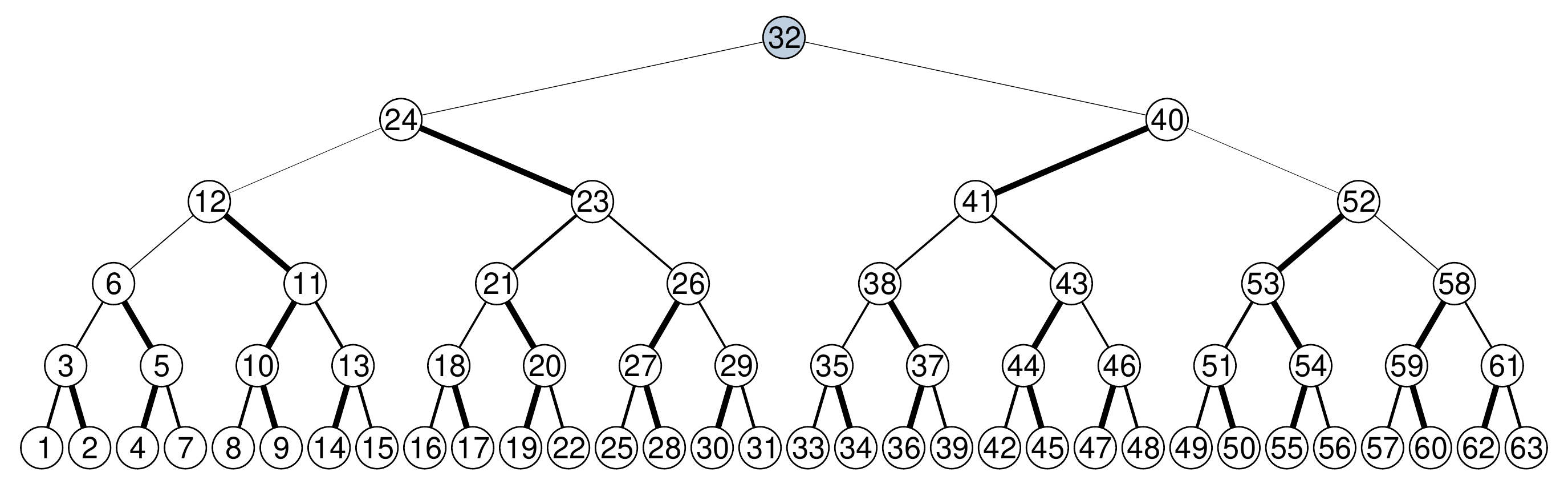}%
  \label{fig:tree-minla}%
}%
\subfloat[{\minbw: \stats{3.629}{4.350}{4.581}{7}}]{%
  \includegraphics[width=0.5\linewidth]{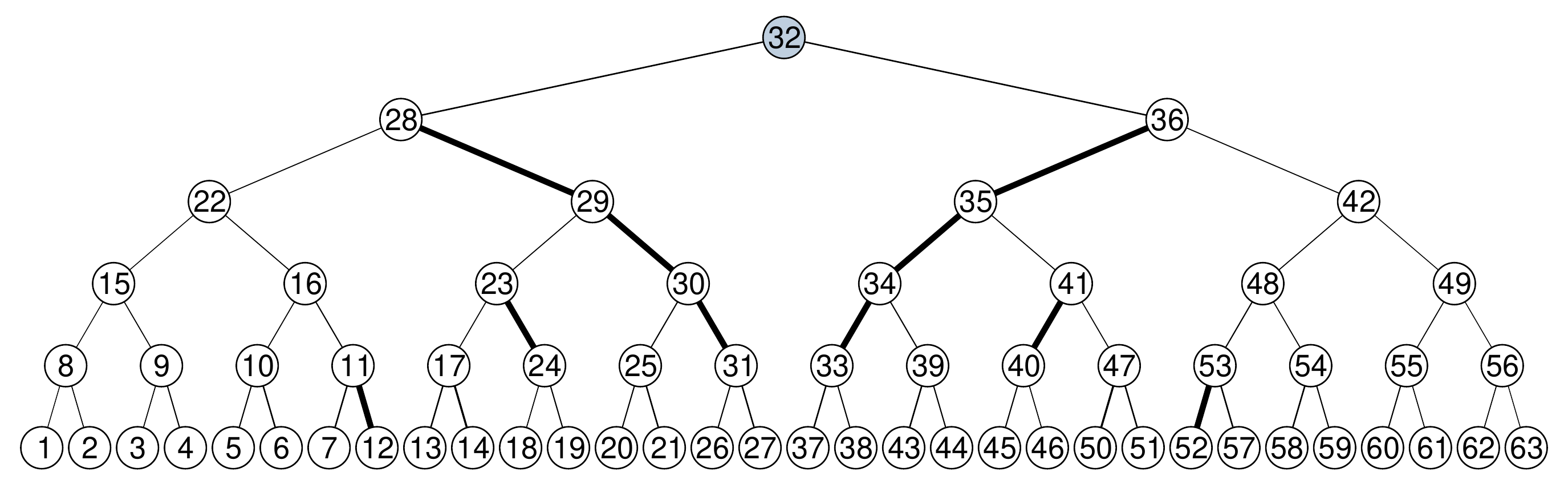}%
  \label{fig:tree-minbw}%
}%
\caption{
  Layouts and locality functionals
  \pwmean[0] (weighted edge product),
  \pwmean[1] (weighted edge sum),
  \pmean[1] (mean edge length),
  and \pmean[\infty] (maximum edge length)
  of a tree with $h = 6$ levels.
  For $h \leq 6$, \minep and \minwep coincide.
  Edges $ij$ are drawn with thickness inversely proportional to length \len[ij].
  Cuts are shown as dashed lines that span the width of subtrees with 3 or more levels.
  Colored vertices are roots of in- (blue) or pre-order (red) subtrees with 2 or more levels.
}
\label{fig:tree-layouts}%
\end{figure*}

%% file: appendix.tex
\clearpage

\appendix

\section{Proofs}
\label{sec:proofs}

\begin{theorem*}
The \minwla layout minimizes \pwmean[1] among all \RLs with cut height $g=1$.
\end{theorem*}
\begin{IEEEproof}
\comment{
We introduce the following notation ot prove that, for binary tree $T$, the \minwla layout minimizes \pmean[1] among all \RLs with cut height $g=1$. 
For trees of height $h$ cut at height $g=1$, let $C_I^{h}$, $C_P^{h}$ be $W$ times average weighted edge length \pwmean[1] when the top subtree is arranged in-order and pre-order, respectively.
}
Recall that the weight of an edge between level $d$ and level $d+1$ is $2^{-d}$.
Observe that when $g=1$, the top subtree $A$ is a single \node $x$, and there are only two bottom subtrees, which we denote as $B_1$ and $B_2$. 
\note{We use subscripts $I$ and $P$ elsewhere. Changed here.}
\comment{
We make no assumptions on the arrangement of the bottom subtrees -- this is what \autoref{thm:order} proves. 
}
Essentially, we have to prove the following for all \RLs with cut height $g=1$. 
\begin{enumerate}
\item
\label{item:allpre}
\pwmean[1] is always minimized by arranging both bottom subtrees $B_{1}$ and $B_{2}$ pre-order.
\item
\label{item:onlytopin}
For any subtree, the optimal in-order arrangement has lower \pwmean[1] than the optimal pre-order arrangement. 
\end{enumerate}
Consider the optimal pre-order arrangement for a subtree of height $h$. Without loss of generality, let $B_1$ be the bottom subtree closest to $A$ in a pre-order arrangement.
We can obtain an in-order arrangement of lower cost by moving the second bottom subtree $B_2$ to the other side of the top subtree $A$, since this moves the root of $B_2$ closer to $x$ without changing the costs of either bottom subtree. Clearly, the optimal in-order arrangement must have lower cost than this in-order arrangement, which proves \autoref{item:onlytopin}. 

Now, consider the optimal in-order arrangement of a bottom subtree, which has height $h-1$. Let this be subtree $B_*$, since the analysis is valid for both bottom subtrees.
By moving one of the bottom subtrees of height $h-2$ to the other side of $B_*$'s root, we convert it to a pre-order arrangement, bringing $B_*$'s root closer to $x$ by $2^{h-2} - 1$. At the same time, if we flip
the order of the \nodes of the bottom subtree we just moved, we 
have increased the length of the edge connecting its root to $B_*$'s root by the same $2^{h-2}-1$. However, since this edge is one level further down the tree, it has a lower weight, and therefore the overall cost has decreased. (Observe that no other edge lengths have changed.) As a result, we can obtain a pre-order arrangement for $B_*$ that is of lower cost than the optimal in-order arrangement for $B_*$. This implies that the optimal pre-order arrangement for $B_*$ has a lower \pwmean[1] value than the optimal in-order arrangement for $B_*$, proving \autoref{item:allpre}.

It is not too difficult to extend this proof for all weight distributions where the weights do not increase from one level to the next. As a result, among all \RLs with cut height $g=1$, 
\minwla optimizes the unweighted measure \pmean[1], and also optimizes \pwmean[1] for the exact weight distribution described in \autoref{eq:edgewts}.
\todo{Finish proof to fix Case 3. Done.}
\end{IEEEproof}

\comment{
\begin{theorem*} 
Proof of \autoref{thm:subtree-in-order}.
\end{theorem*}
\begin{IEEEproof}
Consider an ordering for which $x < y$ and $Y < A < X$.
Let $d = \pos[y] - \pos[x]$ and let $\ell_x$ and $\ell_y$ be the lengths
of the edges connecting $x$ with $X$ and $y$ with $Y$.  Now
swap the positions of $X$ and $Y$ so that $X < A < Y$, and
order the \nodes in $X$ by the corresponding reverse ordering
of \nodes in $Y$, and vice versa.  It is easy to see that this
modification of the layout shortens both $\ell_x$ and $\ell_y$
by $d$ without changing the lengths of all other
edges in the tree. Thus this ordering with $X < A < Y$ reduces all
edge lengths, and will result in an ordering with smaller \PB.
\end{IEEEproof}
}

\begin{theorem*}
For any subtree in a particular branch of the recursion, suppose we fix the internal ordering of the leaves of the top subtree $A$ and the arrangement of all the bottom subtrees in subsequent branches of the recursion. Then, the product of all the edge lengths between the top subtree and the bottom subtrees is minimized by ordering the bottom subtrees in reverse order of that of the parent leaves $L_A$.
\end{theorem*}
\begin{IEEEproof}
Let \pos represent the layout of the top subtree $A$.
Consider two leaves $x$ and $y$ of the top subtree $A$. Let $X$ and $Y$ be bottom subtrees whose parents are
$x$ and $y$, respectively.  We use $x < y$ to mean
$\pos[x] < \pos[y]$; $x < Y$ implies
$x < y \quad \forall y \in Y$; and $X < Y$ implies
$x < y \quad \forall x \in X, y \in Y$.  

Without loss of generality, assume $x < y$.
Consider the case in which the two bottom subtrees $X$ and $Y$ appear
to the same side of the top subtree $A$ containing $x$ and $y$. Without loss of 
generality, assume that $X$ and $Y$ are to the right of $A$, \ie
$y<X$ and $y<Y$. 
We show that if $X < Y$, then there exist operations that
will place $Y$ before $X$ such that the product of the edge lengths is lowered.  Because
the edges $\{x, X\}$ and $\{y, Y\}$ have the same weight, it is
easy to see that the value of this weight does not affect the proof,
and hence we assume that this weight is one. 

Suppose $X < Y$. 
\comment{and consider some arbitrary ordering of the \nodes within
these subtrees.  
}
In \pos, let the distance
between $x$ and $y$ be $d_1$, between $y$ and the first \node of
$X$ be $d_2$, and between the first \nodes of $X$ and $Y$ be $d_3$.
Furthermore, let $r_x$ and $r_y$ denote the distances from the first
\node of $X$ and $Y$ to their roots, such that the lengths of the
edges between $x$ and $y$ and their subtrees are $\ell_x = d_1 + d_2 + r_x$
and $\ell_y = d_2 + d_3 + r_y$, respectively.
(See the top example in \autoref{fig:subtree-ordering}).  

\begin{figure}[tp]
\centering
\includegraphics[width=\columnwidth]{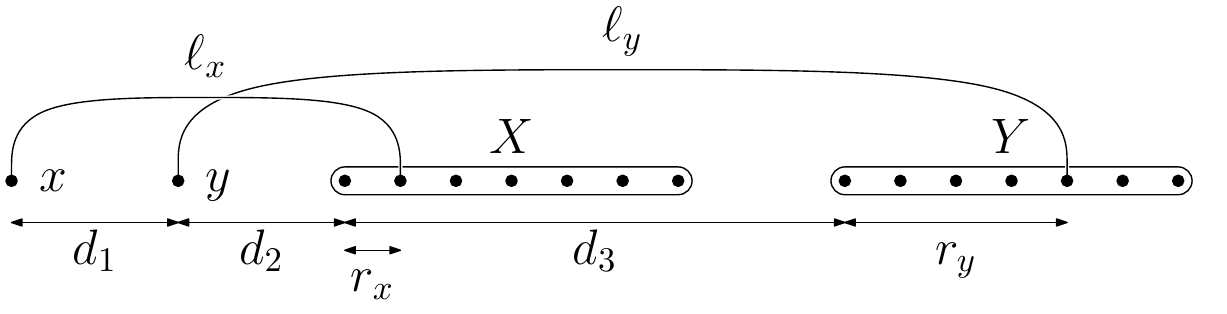}\\[1ex]%
\includegraphics[width=\columnwidth]{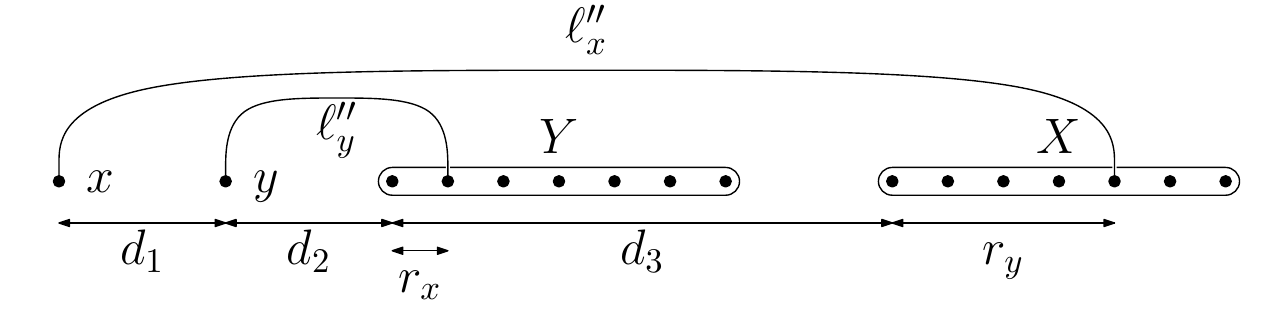}%
\caption{
  The cost is reduced by swapping the positions of $X$ and $Y$.
  The bottom ordering is guaranteed to have a lower
  cost than the top ordering.
}
\label{fig:subtree-ordering}
\end{figure}

The edge product contribution
of the two joining edges is $\ell_x \ell_y$.  
\comment{
while keeping their internal orderings the same.  This results in new edge lengths
$\ell_x' = \ell_x + d_3$ and $\ell_y' = \ell_y - d_3$.  Thus the cost
difference due to the swap is
$\ell_x \ell_y - \ell_x' \ell_y' = d_3 (d_3 + \ell_x - \ell_y)$.
Since $d_3 \geq 1$, this difference is guaranteed positive if
$\ell_x \geq \ell_y$, resulting in a reduced total cost.
If on the other hand $\ell_x < \ell_y$, then we permute 
}
Now consider swapping
the positions of the subtrees $X$ and $Y$ and
permuting the \nodes
within $X$ and $Y$, keeping all other \nodes fixed, so that $X$
takes on the relative ordering of $Y$ and vice versa.  
(See the bottom example in \autoref{fig:subtree-ordering}).  
This
permutation does not affect the internal contributions of $X$
and $Y$ to the cost since the two subtrees have the same structure
(\ie the internal costs of $X$ and $Y$ are interchanged, but their
product remains the same).  However, this operation results in new
edge lengths $\ell_x'' = \ell_y + d_1$ and $\ell_y'' = \ell_x - d_1$
with respect to $x$ and $y$.  The resulting cost difference is
now $\ell_x \ell_y - \ell_x'' \ell_y'' = d_1 (d_1 + \ell_y - \ell_x)$.
And since $d_1 \geq 1$ and by assumption $\ell_y > \ell_x$, the
cost has been reduced.  Consequently, we can always reduce the
cost by enforcing $Y < X$ whenever $x < y$, which completes the
proof.
\end{IEEEproof}

\begin{theorem*}
The \minep layout minimizes \pwmean[0] among all \RLs with cut height $g=1$.
\end{theorem*}
\begin{IEEEproof}
We introduce the following notation ot prove that, for binary tree $T$, the \minep layout minimizes \WEP among all \RLs with cut height $g=1$.
For trees of height $h$ cut at height $g=1$, let $C_I^{h}$, $C_P^{h}$ be $W$ times the logarithm of the weighted edge product functional \pwmean[0] when the top subtree is arranged in-order and pre-order, respectively.
Recall that the weight of an edge between level $d$ and level $d+1$ is $2^{-d}$.
Observe that when $g=1$, the top subtree $A$ is a single \node $x$, and there are only two bottom subtrees, which we denote as $B_1$ and $B_2$. 
\note{We use subscripts $I$ and $P$ elsewhere. Changed here.}
\comment{
We make no assumptions on the arrangement of the bottom subtrees -- this is what \autoref{thm:order} proves. 
}
Essentially, we have to prove the following for all \RLs with cut height $g=1$. 
\begin{enumerate}
\item
\label{item:inord}
If the top subtree $A$ is in-order, then \pwmean[0] is minimized by arranging both bottom subtrees $B_{1}$ and $B_{2}$ pre-order.
\item
\label{item:preord}
If the top subtree $A$ is pre-order, then \pwmean[0] is minimized by arranging the subtree closest to it ($B_{1}$)  pre-order, and the other ($B_{2}$) in-order.
\item
\label{item:topin}
The optimal in-order arrangement has lower \pwmean[0] than the optimal pre-order arrangement. 
\end{enumerate}

Consider the optimal pre-order arrangement with cost $C_P^{h}$, as prescribed by \autoref{item:preord}. We can obtain an in-order arrangement of lower cost by moving the second bottom subtree $B_2$ to the other side of the top subtree $A$, since this moves the root of $B_2$ closer to $x$ without changing the costs of either bottom subtree. Clearly, the optimal in-order arrangement, as prescribed by \autoref{item:inord} must have lower cost than this in-order arrangement, and therefore $C_I^h \leq C_P^h$, which proves \autoref{item:topin}. Therefore, we only need to prove \autoref{item:inord} and \autoref{item:preord}.

First, by inspection, we prove that this is true for $h=\{1,2\}$. When $h = 1$, we trivially have the result, since there are no sub-trees. When $h=2$, \autoref{item:inord} and \autoref{item:preord} are trivial since the subtrees have only one \node (and therefore in-order and pre-order is the same for them). 

Observe that in the in-order case (\autoref{item:inord}), arranging one bottom subtree in-order and the other pre-order will be dominated either by ordering both the bottom subtrees in-order, or both pre-order. Therefore, we only need to compare arranging either bottom subtree pre-order with arranging either bottom subtrees in-order. The optimal choice for the in-order case applies for the bottom subtree closest to the top subtree in the pre-order case (\autoref{item:preord}). Therefore, we only need to need to compare the two choices for arranging the second bottom subtree ($B_2$) in this case.

\note{I just moved this paragraph ``To prove...'', which wasn't useful in the main text.  Need to work it in somewhere. Worked in at top of the proof}

We prove the general case ($h \geq 3$) by induction. Observe that the bottom subtrees $B_1$ and $B_2$ (of height $h-1$) have edges of weight $1/4$ connecting its root to its children; this offset needs to be carefully accounted for. We can calculate the length of the edge between $x$ and the root of the bottom subtrees $B_1$ and $B_2$ based on whether the subtrees are arranged in-order or pre-order. Comparing this with the optimal cost of the bottom subtrees, we need to prove that
\begin{alignat}{2}
\label{inorder}
C_P^{h-1} & \leq C_I^{h-1} + (h-2)\\
\label{preorder}
C_I^{h-1} + \log(2^{h-1}+2^{h-2}-1) & \leq C_P^{h-1} + (h-1)
\end{alignat}
Inequality~\eqref{inorder} proves \autoref{item:inord}, and as a corollary, proves that $C_I^{h} = C_P^{h-1}$. Used together, inequalities~\eqref{inorder} and~\eqref{preorder} prove \autoref{item:preord}, and as a corollary, proves that $C_P^{h} = \frac{1}{2} (C_I^{h-1} + C_P^{h-1} + \log(2^{h-1}+2^{h-2}-1))$. 

Let us assume the induction hypothesis. In other words, for $h-1$, we have
\begin{alignat}{2}
\label{e:inorder}
C_P^{h-2} & \leq C_I^{h-2} + (h-3) \\
\label{e:preorder}
C_I^{h-2} + \log(2^{h-2}+2^{h-3}-1) & \leq C_P^{h-2} + (h-2)
\end{alignat}
which, in turn, imply
\begin{align}
\label{p:inorder}
C_I^{h-1} = C_P^{h-2}\\
\label{p:preorder}
C_P^{h-1} = \frac{1}{2} (C_P^{h-2} + C_I^{h-2} + \log(2^{h-2}+2^{h-3}-1))
\end{align}

Adding $C_P^{h-2}$ to both sides of \eqref{e:preorder}, and multiplying by $\frac{1}{2}$, we have
$\frac{1}{2} (C_P^{h-2} + C_I^{h-2} + \log(2^{h-2}+2^{h-3}-1)) \leq C_P^{h-2} + \frac{1}{2} (h-2)$. 
Substituting in $C^{h-1}$ from \eqref{p:inorder} and~\eqref{p:preorder}, we get
$C_P^{h-1} \leq C_I^{h-1} + \frac{1}{2} (h-2)$, which implies \eqref{inorder}.

We define $f(x) = \log (2^x + 2^{x-1} - 1)$. It is easy to see that for all $x>1$, we have $x < f(x) < x+1$. Therefore, $-2 < f(h-1) - f(h)$, and from \eqref{e:inorder}, we have $C_P^{h-2} < C_I^{h-2} + (h-1) + f(h-1) - f(h)$.
Adding $C_P^{h-2}$ to both sides, and multiplying by $\frac{1}{2}$, we have 
$C_P^{h-2} < \frac{1}{2} (C_P^{h-2} + C_I^{h-2} + f(h-1) + (h-1) - f(h))$. 
Substituting in $C^{h-1}$ from \eqref{p:inorder} and~\eqref{p:preorder}, we get 
$C_I^{h-1} < C_P^{h-1} + \frac{1}{2} ((h-1) + f(h-1) - f(h) - f(h-2))$, which implies 
$C_I^{h-1} < C_P^{h-1} + \frac{1}{2} ((h-1) + f(h-1))$, which in turn implies 
\eqref{preorder}.

It is not too difficult to extend this proof for all weight distributions where the weights do not increase from one level to the next. As a result, 
among all \RLs with cut height $g=1$, \minep optimizes the unweighted measure \pmean[0], and also optimizes \pwmean[0] for the exact weight distribution described in \autoref{eq:edgewts}.
\todo{Finish proof to fix Case 3. Done.}
\end{IEEEproof}

%% file: skeleton.bbl
\begin{thebibliography}{10}
\providecommand{\url}[1]{#1}
\csname url@samestyle\endcsname
\providecommand{\newblock}{\relax}
\providecommand{\bibinfo}[2]{#2}
\providecommand{\BIBentrySTDinterwordspacing}{\spaceskip=0pt\relax}
\providecommand{\BIBentryALTinterwordstretchfactor}{4}
\providecommand{\BIBentryALTinterwordspacing}{\spaceskip=\fontdimen2\font plus
\BIBentryALTinterwordstretchfactor\fontdimen3\font minus
  \fontdimen4\font\relax}
\providecommand{\BIBforeignlanguage}[2]{{%
\expandafter\ifx\csname l@#1\endcsname\relax
\typeout{** WARNING: IEEEtran.bst: No hyphenation pattern has been}%
\typeout{** loaded for the language `#1'. Using the pattern for}%
\typeout{** the default language instead.}%
\else
\language=\csname l@#1\endcsname
\fi
#2}}
\providecommand{\BIBdecl}{\relax}
\BIBdecl

\bibitem{Bayer:Btrees}
R.~Bayer and E.~McCreight, ``Organization and maintenance of large ordered
  indexes,'' \emph{Acta Informatica}, vol.~1, no.~3, pp. 173--189, 1972.

\bibitem{Bender:COStringBT}
M.~A. Bender, M.~Farach-Colton, and B.~C. Kuszmaul, ``Cache-oblivious string
  {B}-trees,'' in \emph{ACM Symposium on Principles of Database Systems}, 2006,
  pp. 233--242.

\bibitem{Prokop}
H.~Prokop, ``Cache-oblivious algorithms,'' Master's thesis, Massachusetts
  Institute of Technology, 1999.

\bibitem{Brodal:COSearch-vEBLayout}
G.~S. Brodal, R.~Fagerberg, and R.~Jacob, ``Cache oblivious search trees via
  binary trees of small height,'' in \emph{ACM-SIAM Symposium on Discrete
  Algorithms}, 2002, pp. 39--48.

\bibitem{Bender05}
M.~A. Bender, E.~D. Demaine, and M.~Farach-Colton, ``Cache-oblivious
  {B}-trees,'' \emph{SIAM J. Comput.}, vol.~35, no.~2, pp. 341--358, Aug. 2005.

\bibitem{Bender:COSearchCost}
M.~A. Bender, G.~S. Brodal, R.~Fagerberg, D.~Ge, S.~He, H.~Hu, J.~Iacono, and
  A.~L\'opez-Ortiz, ``The cost of cache-oblivious searching,''
  \emph{Algorithmica}, vol.~61, no.~2, pp. 463--505, 2011.

\bibitem{Bender:BTDiffKeys}
M.~A. Bender, H.~Hu, and B.~C. Kuszmaul, ``Performance guarantees for {B}-trees
  with different-sized atomic keys,'' in \emph{ACM Symposium on Principles of
  Database Systems}, 2010, pp. 305--316.

\bibitem{Bender:COStreamingBT}
M.~A. Bender, M.~Farach-Colton, J.~T. Fineman, Y.~R. Fogel, B.~C. Kuszmaul, and
  J.~Nelson, ``Cache-oblivious streaming {B}-trees,'' in \emph{ACM Symposium on
  Parallel Algorithms and Architectures}, 2007, pp. 81--92.

\bibitem{Pagh:COHash}
R.~Pagh, Z.~Wei, K.~Yi, and Q.~Zhang, ``Cache-oblivious hashing,'' in \emph{ACM
  Symposium on Principles of Database Systems}, 2010, pp. 297--304.

\bibitem{Yoon05}
S.-E. Yoon, P.~Lindstrom, V.~Pascucci, and D.~Manocha, ``Cache-oblivious mesh
  layouts,'' \emph{ACM Transactions on Graphics}, vol.~24, no.~3, pp. 886--893,
  2005.

\bibitem{Bender:COMesh}
M.~A. Bender, B.~C. Kuszmaul, S.-H. Teng, and K.~Wang, ``Optimal
  cache-oblivious mesh layouts,'' \emph{Theor. Comp. Sys.}, vol.~48, no.~2, pp.
  269--296, Feb. 2011.

\bibitem{Bender:COBloomFilter}
M.~A. Bender, M.~Farach-Colton, R.~Johnson, R.~Kraner, B.~C. Kuszmaul,
  D.~Medjedovic, P.~Montes, P.~Shetty, R.~P. Spillane, and E.~Zadok, ``Don't
  thrash: How to cache your hash on flash,'' \emph{Proc. VLDB Endow.}, vol.~5,
  no.~11, pp. 1627--1637, Jul. 2012.

\bibitem{Yoon06}
S.-E. Yoon and P.~Lindstrom, ``Mesh layouts for block-based caches,''
  \emph{IEEE Transactions on Visualization and Computer Graphics}, vol.~12,
  no.~5, pp. 1213--1220, 2006.

\bibitem{Chung:MinLA}
F.~R.~K. Chung, ``A conjectured minimum valuation tree,'' \emph{SIAM Review},
  vol.~20, no.~3, pp. 601--603, 1978.

\bibitem{Heckmann:MinBW}
R.~Heckmann, R.~Klasing, B.~Monien, and W.~Unger, ``Optimal embedding of
  complete binary trees into lines and grids,'' in \emph{Graph-Theoretic
  Concepts in Computer Science}, ser. Lecture Notes in Computer Science, 1992,
  vol. 570, pp. 25--35.

\bibitem{Safro11}
I.~Safro and B.~Temkin, ``Multiscale approach for the network
  compression-friendly ordering,'' \emph{Journal of Discrete Algorithms},
  vol.~9, no.~2, pp. 190--202, 2011.

\bibitem{Safro09}
I.~Safro, D.~Ron, and A.~Brandt, ``Multilevel algorithms for linear ordering
  problems,'' \emph{Journal of Experimental Algorithmics}, vol.~13, p.~4, 2009.

\bibitem{Cuthill:minbw}
E.~Cuthill and J.~McKee, ``Reducing the bandwidth of sparse symmetric
  matrices,'' in \emph{24th National Conference}, 1969, pp. 157--172.

\end{thebibliography}
